 \documentclass[12pt,preprint]{aastex}

\newcommand{\ye}{Y_{e}}
\newcommand{\nuf}{\phi^{\mathrm{eff}}_{\nu}}
\newcommand{\anuf}{\phi^{\mathrm{eff}}_{\bar{\nu}}}

\begin{document}


\title{Neutrinos and Nucleosynthesis in Gamma-Ray Burst Accretion Disks}


\author{Rebecca Surman\altaffilmark{1,2} and Gail C. McLaughlin\altaffilmark{1}}


\altaffiltext{1}{Department of Physics, North Carolina State University, 
Raleigh, NC  27695}
\altaffiltext{2}{Department of Physics, Union College, Schenectady, NY 
12308}


\begin{abstract} 

We calculate the nuclear composition of matter in accretion disks
surrounding stellar mass black holes as are thought to accompany gamma-ray
bursts (GRBs).  We follow a mass element in the accretion disk starting at
the point of nuclear dissociation and calculate the evolution of the
electron fraction due to electron, positron, electron neutrino and
electron antineutrino captures.  We find that the neutronization of the
disk material by electron capture can be reversed by neutrino interactions
in the inner regions of disks with accretion rates of 1 M$_{\sun}$/s and
higher.  For these cases the inner disk regions are optically thick to
neutrinos, and so to estimate the emitted neutrino fluxes we find the
surface of last scattering for the neutrinos
(the equivalent of the proto-neutron star neutrinosphere) for each optically
thick disk model.  We also estimate the influence of neutrino interactions
on the neutron-to-proton ratio in outflows from GRB accretion disks, and
find it can be significant even when the disk is optically thin to
neutrinos. 

\end{abstract}




\section{Introduction}

It is increasingly thought that the progenitor of a gamma-ray burst may be a
rapidly accreting black hole (see, e.g., \citet{mes02} for a review),
formed from either the collapse of a massive star
\citep{woo93,pac98,mac99,mac01,vie98} or the collision of compact binaries
\citep{eic89,nar92,ruf99,fry99,pac91,ruf01}.  Either case is thought to
result in the formation of a black hole of one to several solar masses
surrounded by a debris torus of similar mass accreting at rates of
fractions to several solar masses per second.  Models of GRB accretion
disks calculated by \citet{mac99} and \citet{nar01} suggest that much of
this disk material is not actually accreted but is lost in vigorous winds. 
If so, GRB's may be important contributors to local, and possibly
galactic, nucleosynthesis, especially for rare species such as $r$-process
nuclei. 

The nuclear composition of the ejected material is determined by processing 
both in the disk and in the outflow.  Within the disk, the nuclei dissociate 
as the accreting material becomes hotter and denser closer to the black hole.  
The nuclear composition is then determined by the forward and reverse 
processes
\begin{mathletters}
\begin{eqnarray}
e^{-} + p & \rightleftarrows & n + \nu_{e} \label{eq:capa} \\
e^{+} + n & \rightleftarrows & p + \bar{\nu}_{e}. \label{eq:capb}
\end{eqnarray}
\end{mathletters}
Electron capture on the liberated protons tends to raise the neutron 
to proton ratio $n/p$, or, equivalently, to reduce the electron 
fraction $\ye = 1/(1+n/p)$.  The extent to which the remaining three 
capture processes mitigate this effect is our primary concern here.

Accretion disk composition has been previously investigated by
\citet{pru02}.  \citet{pru02} dynamically follow the evolution of the
electron fraction within the disk models of \citet{pop99} (hereafter PWF). 
However, they consider only the forward reactions of Eqns.~\ref{eq:capa}
and \ref{eq:capb}, and so are limited to disk models with lower accretion
rates ($\dot{m} \lesssim 1$, where $\dot{m} =
\dot{M}/$(M$_{\sun}$s$^{-1}$)) where neutrinos are likely less important. 
Here, we proceed dynamically as in \citet{pru02} but include electron
neutrino and antineutrino capture in our evaluation of $\ye$. As
anticipated, our solutions for the optically thin PWF disk models do not
differ significantly from \citet{pru02} with the inclusion of neutrinos. 
For accretion rates of 1 M$_{\sun}$/s and higher, though, neutrinos become
quite important in the inner regions of the disk.  For these higher
accretion rates, we shift to the disk models of \citet{dim02} (hereafter
DPN) that explicitly include the effects of neutrino opacity.  In all
cases, we also calculate the equilibrium electron fraction for comparison,
and find it coincides with the dynamical $\ye$ only in the innermost
regions of the disk. 

We also model a simple adiabatic outflow to illustrate the importance of 
neutrino interactions to the evolution of $\ye$ in this material.  We 
find that while the electron and positron capture rates fall off rapidly 
as the material expands and cools, the neutrino rates remain relatively 
constant.  This effect is due entirely to the disk geometry, and so the 
importance of neutrino interactions in the outflow depends quite 
sensitively on the outflow parameters.  Therefore, the neutrino
captures may be important in the outflow, even in the PWF scenarios.

In section~\ref{dcalc} we describe the calculation of the electron fraction 
within the disk.  Section~\ref{result} contains the results of our disk 
calculations and comparisons with previous work.  The simple outflow 
model is discussed in section~\ref{otflmod}.

\section{Disk Calculation} \label{dcalc}

We calculate the evolution of $\ye$ in the disk by following a mass element 
as it moves radially from the point at which all nuclei are dissociated 
(taken here to be $T=10^{10}$ K) to the inner radius of the disk (just 
outside the Schwarzschild radius).  
The method by which our calculation proceeds depends 
on the neutrino opacity of the disk model.  The outer regions of the disk, 
regardless of accretion rate, are optically thin to neutrinos - any neutrinos 
produced via electron or positron capture typically escape the disk without 
interacting.  Here neutrino capture rates are calculated using the neutrino 
fluxes produced by electron/positron capture, and the rates of all four 
reactions are used to find the change in $\ye$.  In the inner regions of the 
disk, particularly if the accretion rate is 1 M$_{\sun}$/s or higher, the 
neutrinos become trapped, and so the calculation of $\ye$ is not as 
straightforward.

The boundary between these zones can be estimated by finding the neutrino 
optical depth as a function of radius within the disk.  At a given radius 
$r$, we estimate the neutrino optical depth as:
\begin{equation}
\label{eq:op}
\tau_{\nu}=\rho \kappa_{\nu} H=H/l_{\nu},
\end{equation} 
where $\rho$ is the baryon density, 
$H$ is the scale height of the accretion disk, the $\kappa_{\nu}$ is the 
neutrino opacity, and $l_{\nu}$ is the neutrino mean free path.  An 
equivalent expression holds for the antineutrino optical depth.  
For both neutrinos and antineutrinos, the opacities and mean free paths 
include charged-current and neutral-current neutrino-nucleon interactions 
and neutrino-electron and neutrino-antineutrino scattering.

\begin{figure}
\plotone{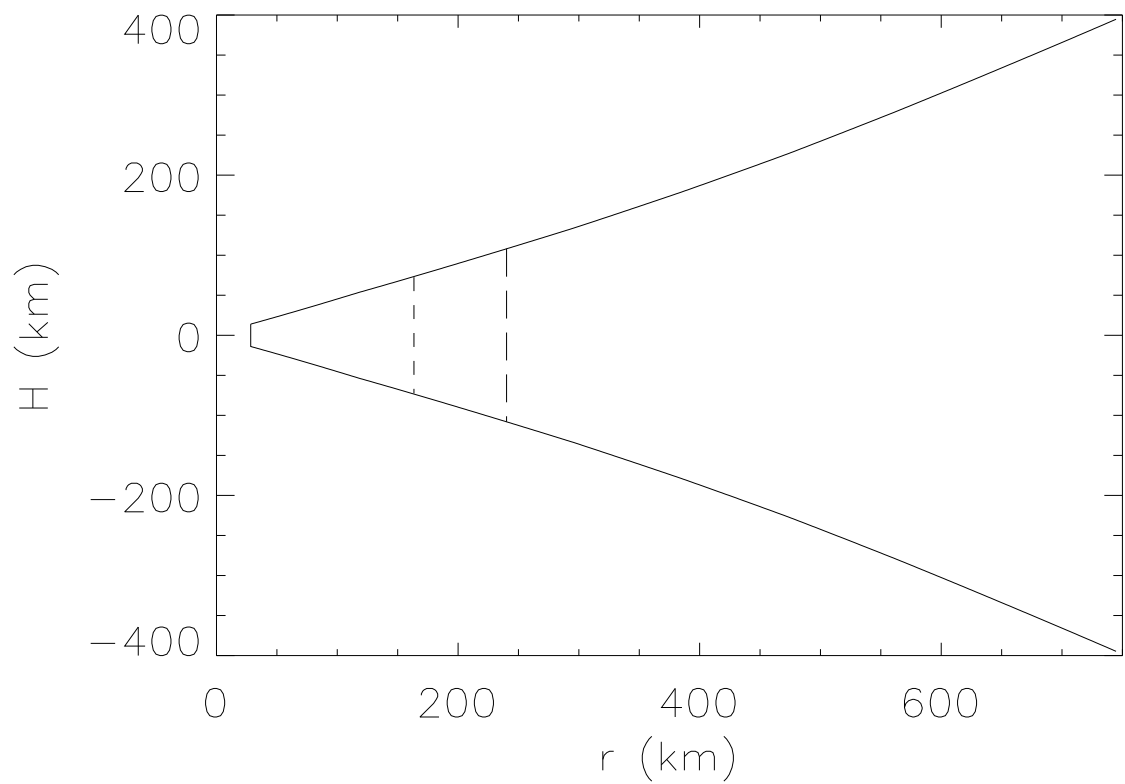}
\caption{Disk profile for the DPN $\dot{m}=10$ disk model.  The scale height 
$H$ of the disk is plotted versus disk radius $r$, from the Schwarzschild 
radius to where the temperature drops below $10^{10}$ K.  The vertical 
long-dashed line marks the radius where the neutrino optical depth $\tau_{\nu}$ 
drops below 2/3, and the short-dashed line marks the equivalent point for 
the antineutrinos. \label{fig:profile}}
\end{figure}

As long as $\tau_{\nu}$, $\tau_{\bar{\nu}} < 2/3$, 
the disk is said to be optically thin and the calculation proceeds as in 
section~\ref{opthin} below.  Where $\tau_{\nu}$, $\tau_{\bar{\nu}} 
> 2/3$, both neutrinos and antineutrinos are trapped and the capture 
reactions of Eqns.~\ref{eq:capa} and \ref{eq:capb} come into equilibrium.  
This is described in section~\ref{opthick}.  Between the optically thin and 
optically thick regions is a zone where the neutrinos are `partially' 
trapped - $\tau_{\nu} > 2/3$ while $\tau_{\bar{\nu}}$ is still $< 2/3$.  
This intermediate zone is treated in section~\ref{partrap}.  
Fig.~\ref{fig:profile} shows a profile of an accretion disk (a DPN 
model with $\dot{m}=10$) illustrating the three zones defined as above.  The 
region to the right of the long-dashed line is optically thin, the region to 
the left of the short-dashed line is optically thick, and partial trapping 
occurs in between. Using this definition, Eq. \ref{eq:op} for
the optical depth, our optical depths agree quite well with DPN,
differing by a maximum of 20\% in the $\dot{m}=10$ case.

We should note that both sets of disk models we employ (PWF, DPN) assume
an electron fraction of 1/2, although in many of the models we find $\ye$
drops far below 1/2.  While we expect the uncertainty this introduces into
our results is small compared to the uncertainties in the disk models
themselves, both types of physics should eventually be included in one
self-consistent calculation.

\subsection{Optically Thin Region} \label{opthin}

The evolution of $\ye$ in the optically thin regions is calculated by
\begin{equation}
\dot{Y}_{e}=v_{r}\frac{d\ye}{dr}=\lambda_{1}-\lambda_{2}\ye \label{eq:yedot}
\end{equation}
where $v_{r}$ is the radial velocity of the disk material, $\lambda_{1} 
= \lambda_{\nu_{e}n} + \lambda_{e^{+}n}$, and $\lambda_{2} = \lambda_{1} 
+ \lambda_{\bar{\nu}_{e}p} + \lambda_{e^{-}p}$. $\lambda_{e^{-}p}$, 
$\lambda_{\nu_{e}n}$, $\lambda_{e^{+}n}$, and $\lambda_{\bar{\nu}_{e}p}$ are 
the rates for the forward and reverse reactions in Eqns.~\ref{eq:capa} and 
\ref{eq:capb}.  Neutron decay is unimportant relative to the short ($\sim$ 
seconds) dynamical timescale and so is not included.  The above expression 
for $\dot{Y}_{e}$ also neglects a (small) general relativistic correction 
and is clearly only valid once all nuclei have dissociated.  
Except where noted, all calculations begin assuming 
$Y_{p} = \ye = 0.5$ at $T=10^{10}$ K.  We follow a mass element by stepping 
through the radial zones $j$, of width $dr_{j}$, of the disk model  
from the point $r_{0}$ where $T\sim 10^{10}$ K to the inner radius of the disk 
$r_{jmax}$, just outside the Schwarzschild radius.  At each zone $j$, we 
explicitly evolve $\ye$ according to Eqn.~\ref{eq:yedot} above, so that 
\begin{equation}
Y_{e,j} = Y_{e,j-1} + \Delta \ye = Y_{e,j-1} + 
(\lambda_{1}-\lambda_{2}Y_{e,j-1})dr_{j}/v_{r,j}. \label{eq:ye}
\end{equation}

The electron and positron capture rates 
$\lambda_{e^{-}p}$ and $\lambda_{e^{+}n}$ are given by
\begin{eqnarray}
\lambda_{e^{-}p} & = & \int_{\Delta_{np}}^{\infty} 
\sigma_{e}(E_{e},-\Delta_{np}) c f_{e}(E_{e},\mu_{e}) dE_{e}\\
\lambda_{e^{+}n} & = & \int_{m_{e}c^{2}}^{\infty} 
\sigma_{e}(E_{e},\Delta_{np}) c f_{e}(E_{e},-\mu_{e}) dE_{e}\\
\end{eqnarray}
where 
\begin{equation}
\sigma_{e}(E_{e},Q) = \frac{1}{64 \pi} \left (\frac{g_{W}}{M_{W}c^{2}}\right)^{4} 
(\hbar c)^{2} (c_{V}^{2} + 3c_{A}^{2}) \cos^{2}\theta_{C} (E_{e} + Q)^{2}
\label{eq:sige}
\end{equation}
and
\begin{equation}
f_{e}(E_{e},\mu_{e}) = \frac{1}{\pi^{2}(\hbar c)^{3}}
\frac{p_{e}c E_{e}}{e^{(E_{e}-\mu_{e})/kT}+1}.
\label{eq:fe}
\end{equation}
Here $\Delta_{np} \approx 1.293$ MeV is the neutron-proton mass
difference, $\mu_{e}$ is the electron chemical potential, $m_{e}c^{2}
\approx 0.511$ MeV is the electron mass, $g_{W}\approx 0.65$ is the
dimensionless weak coupling constant, $M_{W}c^{2} \approx 80.9$ GeV is the
mass of the $W$ boson, $c_{V} \approx 1$ and $c_{A}\approx 1.26$ are the
vector and axial vector couplings, and $\theta_{C}\approx 13\degr$ is
the Cabibbo angle.  The electron chemical potential is set by the
requirement $n_{e^{-}} - n_{e^{+}} = \rho N_{A} \ye$, where
$n_{e^{-}}$,$n_{e^{+}}$ are the Fermi-Dirac number densities for the
electrons and positrons, respectively, $\rho$ is the baryon density, and
$N_{A}$ is Avogadro's number. 

The neutrino and antineutrino capture rates $\lambda_{\nu_{e}n}$ and 
$\lambda_{\bar{\nu}_{e}p}$ are given by
\begin{eqnarray}
\lambda_{\nu_{e}n} & = & a \int^{\infty}_{0} 
(E+\Delta_{np})^{2} \nuf dE \\
\lambda_{\bar{\nu}_{e}p} & = & a \int^{\infty}_{\Delta_{np}+m_{e}c^{2}} 
(E-\Delta_{np})^{2} \anuf dE,
\end{eqnarray}
where 
\begin{equation}
a=\frac{(\hbar c)^{2}}{32 \pi} \left(\frac{g_{W}}{M_{W}c^{2}}\right)^{4}
(c_{V}^{2} + 3c_{A}^{2})
\end{equation}
and $\nuf$, $\anuf$ are the effective neutrino and 
antineutrino fluxes in units of 1/(cm$^{2}\cdot$s$\cdot$keV).  
Note that here as elsewhere in the paper, we have neglected general
and special relativistic effects, c.f. \citet{pfc}.

The effective fluxes at each radius within the disk include not only the
neutrinos and antineutrinos produced by electron and positron capture at
that radius but contributions from neutrinos and antineutrinos produced
throughout the disk.  This introduces a complication in our overall disk
calculation, as the effective neutrino flux in the outer radial zones
depends on the flux emitted from the inner zones, which in turn is
sensitive to the yet uncalculated local $\ye$.  We therefore require
several iterations of our overall disk calculation.  In the first
iteration, we dynamically solve for $\ye$ throughout the disk according to
Eqn.~\ref{eq:ye} with the neutrino capture rates $\lambda_{\nu_{e}n}$ and
$\lambda_{\bar{\nu}_{e}p}$ set to zero. At each zone $j$, we calculate the
neutrino and antineutrino fluxes \textit{emitted} from electron and
positron capture within that zone by
\begin{eqnarray}
\phi_{\nu,j} & = & \sigma_{e}(E_{e},-\Delta_{np}) c f_{e}(E_{e},\mu_{e}) 
n_{p} dr_{j} \label{eq:nuj}\\
\phi_{\bar{\nu},j} & = & \sigma_{e}(E_{e},\Delta_{np}) c f_{e}(E_{e},-\mu_{e})  
n_{n} dr_{j}, \label{eq:anuj}
\end{eqnarray}
where $\phi_{\nu,j}$ and $\phi_{\bar{\nu},j}$ give the number of neutrinos
and antineutrinos emitted per second per keV per unit area of the emitting
region, $dr_{j}$ is the zone thickness, $\sigma_{e}$ and $f_{e}$ are
defined in Eqns.~\ref{eq:sige} and \ref{eq:fe} above, and $n_{p}$ and
$n_{n}$ are the proton and neutron number densities, respectively. In the
second and subsequent iterations, we repeat the evolution of $\ye$ through
the disk including $\lambda_{\nu_{e}n}$ and $\lambda_{\bar{\nu}_{e}p}$,
with the effective neutrino fluxes at each zone determined by the emitted
neutrino fluxes of the previous iteration (described below). We continue
to iterate the disk calculation until $\ye$ throughout the disk differs
from the previous iteration by an average of less than $1\%$. 

The calculation of the effective neutrino fluxes at each zone requires 
some care, particularly since the disk geometry seems to necessitate a 
fully three-dimensional calculation. In general, each disk zone $j$ is 
a thin cylindrical shell of radius $r_{j}$, thickness $dr_{j}$, and 
total height 
equal to twice 
the local scale height $H_{j}$ of the disk.  The neutrinos produced  
in each zone are taken to be emitted from the surface area of this shell, 
and so the emitted fluxes $\phi_{\nu,j}$ and $\phi_{\bar{\nu},j}$ are in 
actuality functions of height in the disk, since we expect the temperature 
and density, and therefore the electron and positron capture rates, to 
drop with height.  This fact complicates the evaluation of the 
effective neutrino flux, since (nominally) the effective flux in zone 
$i$ is 
\begin{equation} 
\nuf = \sum_{j=0}^{jmax} \int \phi_{\nu,j} 
d\Omega_{i,j}/4\pi, \label{eq:phij}
\end{equation}
where $d\Omega_{i,j}$ is the solid angle subtended by zone $j$ as viewed 
from zone $i$.  Instead of embarking on the involved and numerically 
expensive exact evaluation of this expression, we take the emitted neutrino 
flux from each zone to be a height-adjusted constant, so that $\phi_{\nu,j}$ is 
decoupled from the solid angles $d\Omega_{i,j}$. The solid 
angles, depicted in Fig.~\ref{fig:solidang}, can then be evaluated 
separately, as described in the appendix.

In order to estimate the height adjustment for the emitted neutrino flux,
we first model the thermodynamics of the disk as a function of height. We
simply take the disk to be an adiabatic gas in hydrostatic equilibrium,
where the total pressure is the sum of gas and radiation pressures.  We
use the temperature and density profiles thus generated to estimate the
emitted neutrino flux as a function of height, $\phi_{\nu,j}(z)$, by
applying Eqns.~\ref{eq:nuj} and \ref{eq:anuj} at each height $z$.  The
correction factor is then
\begin{equation}
\mathrm{height\ correction} = h_{j} = \frac{\int \phi_{\nu,j}(z) dz}
{\phi_{\nu,j}\cdot H_{j}}, \label{eq:h}
\end{equation}
where $\phi_{\nu,j}$ is the emitted neutrino flux at $z=0$.  An equivalent 
expression holds for the antineutrinos.

The effective fluxes $\nuf$, $\anuf$ require yet another correction.  In
principle, a neutrino diffusion calculation is required here since as we
approach the optically thick part of the disk, the neutrinos will start to
scatter.  The transition between the optically thin and thick regions is
much less sharp than in the protoneutron star of the core collapse
supernova.  While not undertaking the full-blown calculation, we still
approximate this scattering by including an extinction correction. As the
neutrino flux emitted from zone $j$ travels to zone $i$, it will be at
least partially extinguished by scattering and captures in the intervening
disk material.  This effect is quite difficult to account for exactly, in
particular because the neutrinos can follow any number of possible paths
from zone $j$ to zone $i$.  We find a minimum extinction correction by
assuming the neutrinos follow the minimum path from $j$ to $i$, so that
\begin{equation}
\mathrm{extinction\ correction} = x_{i,j} = \prod_{k=j}^{i} 
\exp(-dr_{k}/l^{\mathrm{cap}}_{k}), \label{eq:ex}
\end{equation}
where $dr_{k}$ is the width of zone $k$ and $l^{\mathrm{cap}}_{k}$ is the 
neutrino mean free path for captures only in zone $k$.  We expect this 
to be a reasonable 
approximation as long as $r_{i} > r_{j}$.  For $r_{i} < r_{j}$, the maximum  
possible path length between $j$ and $i$ can greatly exceed the minimum, 
and so Eqn.~\ref{eq:ex} will be an underestimate.  However, since the neutrino 
flux typically rises sharply in the interior of the disk, only a small 
fraction of the effective neutrino flux in a given zone comes from larger 
radii and so we don't expect the underestimate to be a problem.

The effective neutrino flux of zone $i$ is therefore
\begin{equation} 
\nuf = \sum_{j=0}^{jmax} h_{j} x_{i,j} \phi_{\nu,j} 
\int d\Omega_{i,j}/4\pi, \label{eq:totalflux}
\end{equation}
where $h_{j}$ and $x_{i,j}$ are given by Eqns.~\ref{eq:h} and \ref{eq:ex}, 
respectively, and $\int d\Omega_{i,j}/4\pi$ is calculated as described 
in the appendix.  The effective antineutrino flux is calculated in the 
same fashion.  If the disk is entirely optically thin ($\tau_{\nu}$,  
$\tau_{\bar{\nu}} < 2/3$ everywhere), then in the second iteration and 
beyond we find the effective neutrino flux as above for each zone, and 
dynamically calculate $\ye$ as previously described.  If not, we treat only 
the optically thin part of the disk in this fashion and switch our 
approach to that described below once trapping sets in.

\subsection{Partial Trapping} \label{partrap}

Once $\tau_{\nu}$ drops below 2/3, the neutrinos begin to be trapped
vertically within the disk.  Again, this would best be described using a
one- or two-dimensional Boltzmann neutrino transport calculation,
particularly since $\tau_{\nu}$ changes rather slowly with radius. 
However, we don't believe this level of sophistication is currently
warranted given the still-large uncertainties in the disk models
themselves.  Instead, once the neutrinos become trapped as defined above,
we assume the neutrinos become thermalized and the forward and reverse
reactions in Eqn.~\ref{eq:capa} come into equilibrium.  Therefore we
replace the effective neutrino flux for these zones with a flux calculated
from a simple Fermi-Dirac distribution,
\begin{equation}
\phi_{\nu}^{FD} = \frac{g_{\nu} c}{2 \pi^{2} (\hbar c)^{3}}  
\frac{E_{\nu}^{2}}{e^{(E_{\nu}-\mu_{\nu})/kT_{\nu}}+1}. \label{eq:FDflux}
\end{equation}

Here $g_{\nu}$ is one and $\mu_{\nu}$ is calculated from the equilibrium
condition $\mu_{\nu} + \mu_{n} = \mu_{e} + \mu_{p}$, or $\mu_{\nu} =
\mu_{e} - \hat{\mu}$, where $\hat{\mu} = \mu_{n}-\mu_{p}$ is calculated
using the \citet{lat91} equation of state.   $T_{\nu}$ is taken to be
the local temperature of the disk material.  It is corrected for height in a
manner similar to Eqn.~\ref{eq:h}, except here the integral is over one mean
free path instead of the entire height of the disk.  Aside from this one
change, the calculation proceeds as in Section~\ref{opthin} above. 

\subsection{Optically Thick Region} \label{opthick}

Once $\tau_{\bar{\nu}}$ drops below 2/3 as well, both neutrino species are
considered trapped and we begin solving for $\ye$ assuming the reactions
of Eqns.~\ref{eq:capa} and \ref{eq:capb} are in equilibrium.  In addition,
we assume lepton number $Y_{L} = Y_{e} + Y_{\nu}$, where $Y_{\nu} =
(n_{\nu_{e}} - n_{\bar{\nu}_{e}})/\rho N_{A}$, leaks very slowly out of
the optically thick region so that $Y_{L}$ can effectively be considered a
constant.  The value of $Y_{L}$ is set to $\ye$ at the edge of the
optically thick region, and we search for values of $\ye$ and $Y_{\nu}$ at
every radial zone $j$ that satisfy
\begin{eqnarray}
\mu_{e^{-}} + \mu_{e^{+}} & = & 0 \\
\mu_{\nu} + \mu_{\bar{\nu}} & = & 0 \\
\mu_{\nu} + \mu_{n} & = & \mu_{e} + \mu_{p},
\end{eqnarray}
since the temperature here is well over an MeV.
Again, $\hat{\mu} = \mu_{n} - \mu_{p}$ is calculated as in \citet{lat91}, and 
the number densities $n_{e^{-}}, n_{e^{+}}, n_{\nu}$, and $n_{\bar{\nu}}$ 
are taken to be Fermi-Dirac with chemical potentials as above and 
temperatures equal to the disk temperature in that zone.  Solving this 
system of equations proceeds as follows: for each zone $j$, we guess a $\ye$, 
find $\mu_{e}$ by inverting $n_{e^{-}} - n_{e^{+}}= \rho N_{A} \ye$, 
calculate $\hat{\mu}$ and solve for $\mu_{\nu}$, find $n_{\nu}$ and 
$n_{\bar{\nu}}$, and check to see if the resulting $Y_{\nu}$ plus $\ye$ keeps 
$Y_{L}$ constant. Such a procedure is similar to that described in 
\citet{belo}, except that in that calculation there is no way to estimate
the appropriate value of $Y_L$ at the edge of the optically thick region, 
since the system was not considered dynamically.

In addition to finding $\ye$, we need to estimate the neutrino flux
emitted from the optically thick region.  In doing so we are guided by the
example of the proto-neutron star (PNS).  In the PNS case, the neutrino
luminosities can be found from the rate at which gravitational binding
energy is released, and the neutrino temperatures can be estimated by
determining where the neutrinos decouple from the baryons and electrons
within the protoneutron star.  Here we cannot follow this prescription for
finding the neutrino luminosities as we have no means to determine how
much energy is lost into the black hole; instead, we calculate the
neutrino and antineutrino fluxes from thermal Fermi-Dirac distributions
(Eqn.~\ref{eq:FDflux}) with one correction.  We add to these fluxes the
neutrinos emitted from the optically thin portions of the disk into the
optically thick region. This small correction is necessary to ensure that
these neutrinos don't simply disappear into the optically thick region. 
This has a negligible effect on the result and if
we had a perfect blackbody, it wouldn't be needed.  Contrary to the
case of the protoneutron star, the boundary between trapping and no-trapping
is not at all sharp.  We in effect artificially draw such a boundary,
and therefore use this and the extinction correction to try to correct for
this sharp boundary approximation.
We should note that the above prescription for the neutrino and
antineutrino fluxes gives values of the same order of magnitude as those
calculated from energy considerations assuming no energy is lost into the
black hole. 

In order to follow the PNS example to evaluate the neutrino temperatures,
we need to translate the concept of a neutrinosphere to the disk geometry. 
In general, the last scattering surface for a sphere can be defined as the
radius $R_{\nu}$ at which the neutrino optical depth $\tau_{\nu}$ becomes
2/3: 
\begin{equation}
\tau_{\nu} = \int_{R_{\nu}}^{\infty} \kappa_{\nu}(E_{\nu},r) \rho(r) dr 
= \frac{2}{3}, \label{eq:nusphere}
\end{equation}
where $\kappa_{\nu}$ is the neutrino opacity and $\rho$ is the baryon density.
It makes little sense to apply this expression to the plane of the disk 
(integrating from the outer edge of the disk inward), as the neutrinos that, 
by the above definition, are `trapped' horizontally can easily escape the 
disk vertically.  Instead, we adopt Eqn.~\ref{eq:nusphere} to calculate where 
the neutrinos decouple vertically as a function of disk radius.  We take the 
vertical disk structure determined as described in section~\ref{opthin}, and 
calculate the height $h_{\nu}$ at which the following expression is satisfied:
\begin{equation}
\tau_{\nu}=\int_{h_{\nu}}^{h_{max}} \frac{1}{l_{\nu}(r)} dr = 
{2 \over 3}, \label{eq:nuh}
\end{equation}
where $h_{max}>>H$ and the neutrino opacities $\kappa_{\nu}$ are taken to
be equal to $(l_{\nu} \rho)^{-1}$, where $l_{\nu}$ is the neutrino mean
free path, here averaged over energy.  

\begin{figure}
\plotone{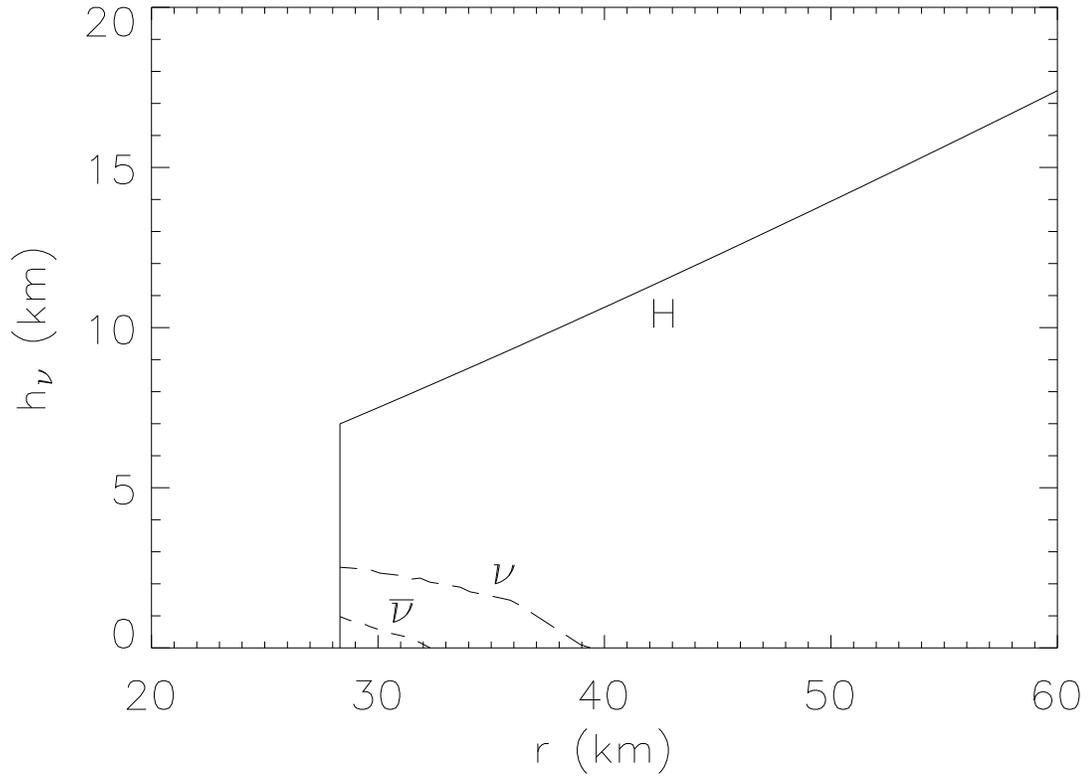}
\caption{Decoupling heights $h_{\nu}$ for neutrinos (long dashes) and 
antineutrinos (short dashes) versus radius in the DPN disk model $\dot{m}=1.0$. 
The solid line marks the scale height $H$ of the disk. \label{fig:nusph1}}
\end{figure}

\begin{figure}
\plotone{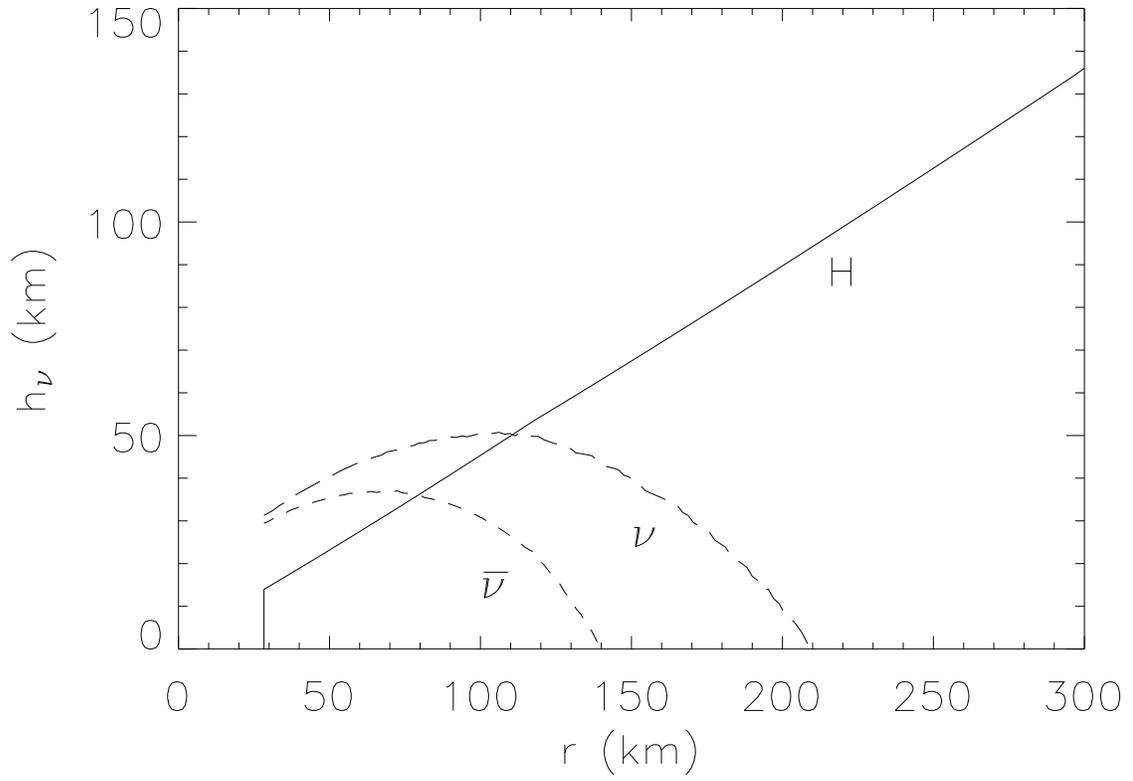}
\caption{Same as Fig.~\ref{fig:nusph1} for the DPN disk model $\dot{m}=10$. 
\label{fig:nusph}}
\end{figure}

The shapes of the neutrino decoupling surfaces thus calculated are shown
in Figs.~\ref{fig:nusph1} and \ref{fig:nusph}.  The long dashed lines show
the decoupling heights $h_{\nu}$ as a function of radius for the
neutrinos; the short dashed lines are for the antineutrinos.  The scale
heights $H$ (solid line) are plotted for comparison.  Contrary to the PNS
case, these are not spherical.  As Fig.~\ref{fig:nusph1} shows, the
decoupling heights $h_{\nu}$ in the DPN $\dot{m}=1$ model steadily
increase with decreasing radius, resulting in wedge-shaped decoupling
surfaces.  In the DPN $\dot{m}=10$ model, the optically thick region is so
large that the decoupling surfaces are additionally shaped by the physical
height of the disk.  Within the radial extent of the optically thick
region the scale height $H$ decreases appreciably.  Therefore while
$h_{\nu}/H$ and the emerging neutrino and antineutrino temperatures
continue to increase with decreasing radius, the decoupling heights
$h_{\nu}$ actually decrease, resulting in the rounded 'neutrinosurfaces'
seen in Fig.~\ref{fig:nusph}. 

A further simplification is made to facilitate the calculation of the
neutrino and antineutrino fluxes emitted horizontally into the outer disk. 
Instead of the complicated shapes shown in 
Figs.~\ref{fig:nusph1}~and~\ref{fig:nusph}, 
we take
the `neutrinosurface' to be a cylinder, with radius $r_{\nu}$ equal to the
outer tip of the `neutrinosurface' (where $h_{\nu} \rightarrow 0$) and
height equal to the maximum decoupling height.  The temperature and
chemical potential of the emerging neutrinos are taken to be their local
values at $r_{\nu}$.  The temperatures that we find range from
around $T_\nu = 2.5 - 4.5 \, {\rm MeV}$ and 
$T_{\bar{\nu}} = 3.6 - 5.1 \, {\rm MeV}$
depending on the model.   
The neutrino and antineutrino fluxes found in this
way are included in the effective neutrino flux calculations described in
section~\ref{opthin}. 

\section{Results - $\ye$ in Disk} \label{result}

The results from our calculations of the electron fraction $\ye$ in the
disk models of PWF and DPN are summarized in Table 1, which contains the
final values of $\ye$ in the innermost radial zones of each disk model. In
this table $\dot{m}$ is the mass accretion rate, {\it a} is the black hole spin
parameter, and $\alpha$ is the viscosity. 

For disks with accretion rates $\dot{m} \lesssim 0.1$, the evolution of
the electron fraction is dominated by electron capture.  For these low
accretion rates the entire disk is typically optically thin to neutrinos,
and so neutrino interactions play a limited role in the disk.  For disks
with higher accretion rates, $\dot{m} \gtrsim 1$, $\ye$ is set by all four
capture reactions.  In these disks, electron capture initially drives
$\ye$ to very low values ($\ye \lesssim 0.1$).  As the mass element
spirals inward, the material becomes optically thick to neutrinos and
neutrino interactions become increasingly important.  Neutrino and
positron capture significantly raise $\ye$ in the inner disk to the final
values in Table 1. 

\subsection{$\dot{m}<1$} \label{thinresults}

\begin{figure}
\plotone{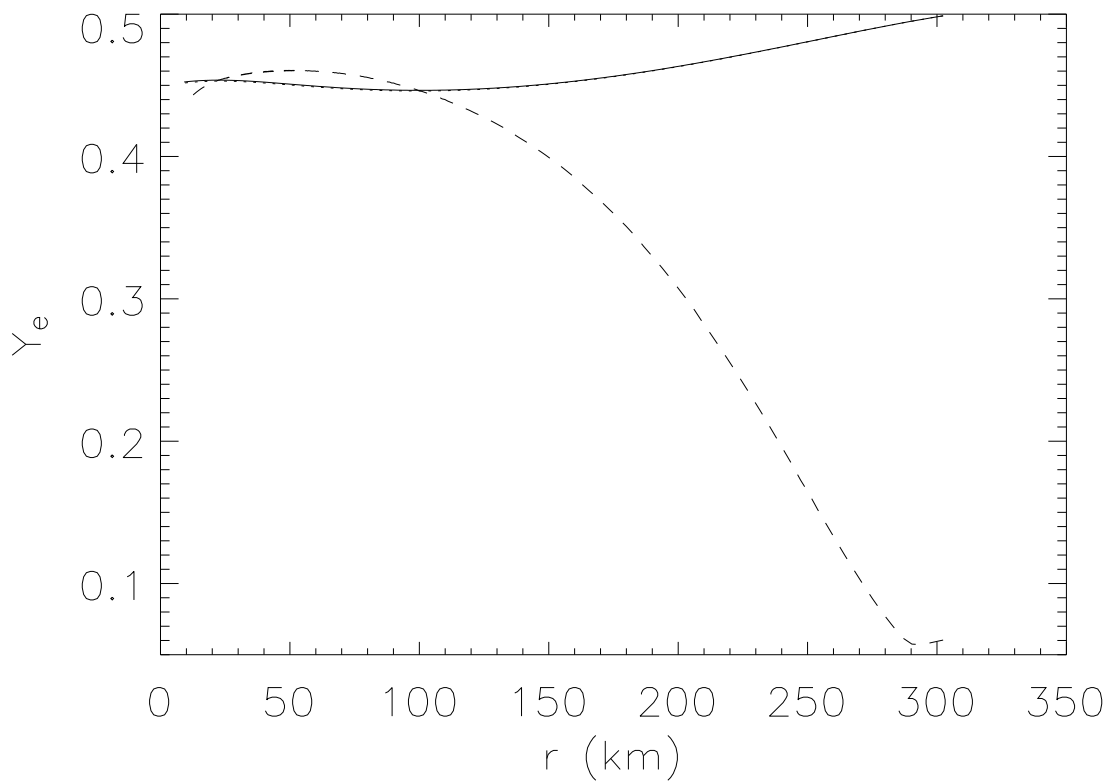}
\caption{Electron fraction $\ye$ as a function of radius for the 
PWF models with $\dot{m}=0.1$, alpha viscosity $\alpha=0.1$, and black 
hole spin parameter $a=0$.  The solid line shows $\ye$ from our full 
calculation, while the dotted line is our calculation without neutrino 
interactions.  The dashed line is the $Y_e$ that would obtain, were the 
system in weak equilibrium.
\label{fig:pwf5}}
\end{figure}

\begin{figure}
\plotone{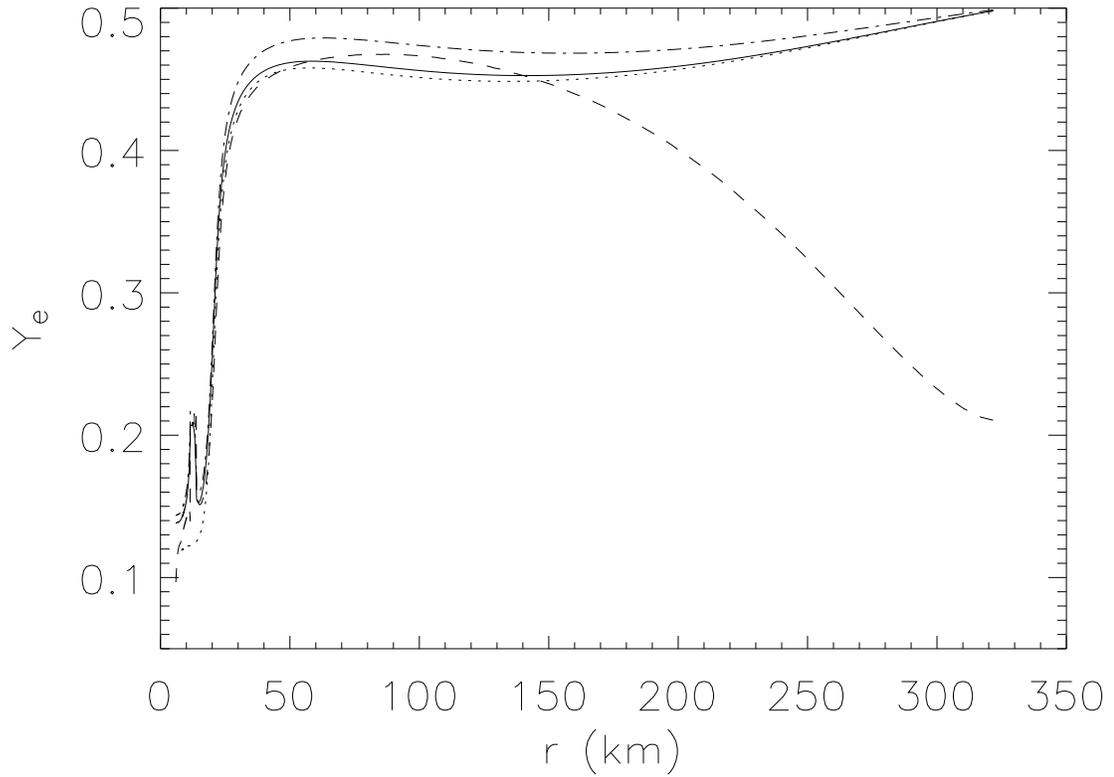}
\caption{Identical to Fig.~\ref{fig:pwf5} for the PWF disk model with 
$\dot{m}=0.1$, $\alpha=0.1$, and $a=0.95$.  Again the solid line is
the full calculation and the dotted is the calculation without neutrinos.
The rise at the center of the disk is due to neutrino trapping.
The additional dot-dashed line 
shows our calculation with just the antineutrinos turned off.
\label{fig:pwf8}}
\end{figure}

Figs.~\ref{fig:pwf5} and \ref{fig:pwf8} show two representative
calculations of the evolution of $\ye$ in optically thin accretion disks,
using PWF disk models with $\dot{m}=0.1$, alpha viscosity $\alpha=0.1$,
and black hole spin parameter $a=0$ (Fig.~\ref{fig:pwf5}) and $a=0.95$
(Fig.~\ref{fig:pwf8}).  For each model, the solid line is our full
calculation and the dotted line is our calculation without the neutrino
interactions.  The dashed lines show for comparison the equilibrium
electron fractions
\begin{equation}
\ye^{\mathrm{eq}} = \frac{\lambda_{1}}{\lambda_{2}},
\end{equation}
where $\lambda_{1}$ and $\lambda_{2}$ are defined as in Eqn.~\ref{eq:yedot}. 

The calculated electron fraction shown in Fig.~\ref{fig:pwf5} is
particularly representative of the PWF $\dot{m}=0.01$, 0.1 models.  The
temperature and density are relatively low so we see only some electron
capture and a small drop in $\ye$, halted by positron capture within $r
\sim 100$ km.  The capture rates are too slow for $\ye$ to equilibrate,
and neutrinos have almost no effect, as stated in \citet{pru02}.  Models
with lower alpha viscosity $\alpha < 0.1$ have higher densities and so
more electron capture (and a correspondingly lower $\ye$), but neutrinos
are similarly unimportant.  However, in the high spin ($a>0$) models the
neutrino capture reactions begin to play a role.  In these models the
portion of the disk close to the Schwarzschild radius is hotter and many
times denser than the equivalent disk with $a=0$. In these conditions the
neutrinos may even become trapped. 

An example of this is shown in Fig.~\ref{fig:pwf8}.  As in
Fig.~\ref{fig:pwf5}, $\ye$ drops initially due to electron capture, and
then rises slightly as positron capture becomes more important.  The steep
drop in $\ye$ beginning at $r \sim 30$ km is due to a rapid increase in
density (raising the electron capture rate) combined with a small dip in
temperature (dropping the positron density and therefore the positron
capture rate).  Once the temperature again begins to rise, positron
capture rebounds and $\ye$ levels off.  Within a small portion ($10 < r <
14$ km) of this very dense and hot region the neutrinos become trapped. 
The corresponding jump in the neutrino capture rate raises $\ye$ within
this narrow zone.  In this small region the crossing time is
comparable to neutrino capture time, about half a millisecond.  However,
this is not a concern for our calculation, since we do not assume weak
equilibrium.  The fact that the neutrinos (not the antineutrinos) 
are trapped means that their flux and spectra are calculated assuming
thermal equilibrium (see section 2.2),
but $Y_e$ is calculated in this region by integration of the
weak rates.

\begin{figure}
\plotone{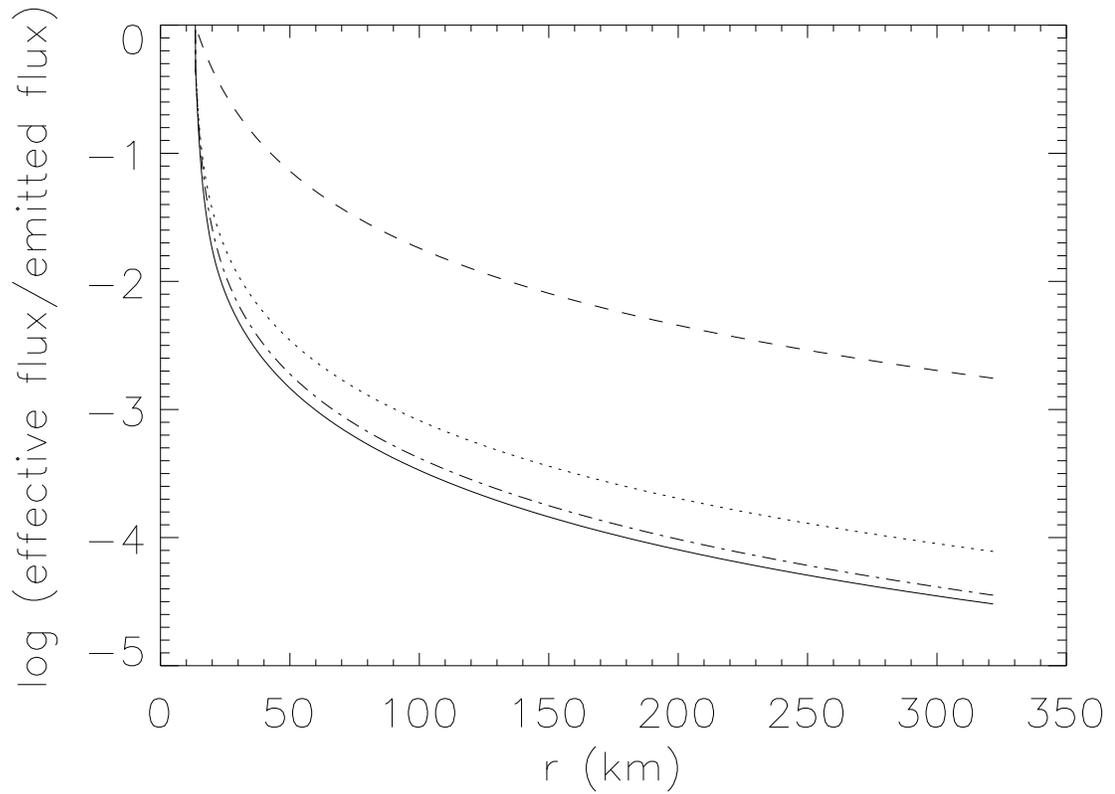}
\caption{Effective neutrino fluxes from the optically thick region divided
by the net flux emitted by the optically thick region, for the PWF disk
model with $\dot{m}=0.1$, $\alpha=0.1$, and $a=0.95$.  The dot-dashed line
is for the antineutrino flux and the solid and dotted lines are for the
neutrino flux calculated with and without the extinction correction,
respectively.  The dashed line shows $r^{-2}$ for comparison. 
\label{fig:geo}}
\end{figure}

Neutrinos have comparatively little effect elsewhere within the disk, as
can be seen by comparing the full calculation in Fig.~\ref{fig:pwf8}
(solid line) to the calculation without neutrino interactions (dotted
line).  It is not immediately apparent that this should be the case, as it
may seem that the high flux from the optically thick region should produce
appreciable neutrino capture throughout the disk.  This does not happen
because the effective neutrino flux from the optically thick region (or,
in fact, from any region in the disk) is rapidly extinguished via two
separate effects.  The first is the extinction correction from section
\ref{opthin}.  The neutrino opacities in the regions just outside of the
optically thick zone are still quite high, so a portion of the flux is
lost to neutrino capture.  The second, and most important, effect is
entirely geometric.  At any radius, much of the neutrino and antineutrino
flux will leave the top or bottom of the disk without interacting with
disk material.  As a result, the effective flux drops off much faster
than, for example, in the proto-neutron star case, where the flux falls as
$r^{-2}$.  This is illustrated in Fig.~\ref{fig:geo}, which shows the
effective neutrino flux from the optically thick region over the net flux
emitted from that region as a function of $r$ (solid line).  The dashed
line is the same quantity with the extinction correction removed, so that
only the geometric effect is reducing the flux.  The dotted line shows
$r^{-2}$ for comparison. 

In addition, in this model the neutrino and antineutrino fluxes
effectively negate each other's influence in the outer disk.  Even though
the neutrino flux from the center of the disk is much higher than the
antineutrino flux, the neutrino opacities are everywhere larger than for
the antineutrinos, leading to greater extinction of the neutrino flux as
it passes through the disk.  This effect is also shown in
Fig.~\ref{fig:geo}, where the effective flux over the emitted flux for
antineutrinos is given by the dot-dashed line.  As a result of this, the
effective neutrino flux for $r > 25$ km is only slightly higher than the
effective antineutrino flux, and the net impact of the neutrinos is small. 
The extent that this effect influences $\ye$ is illustrated by the
dot-dashed line in Fig.~\ref{fig:pwf8}.  It shows a calculation where the
antineutrino interactions only are removed and the neutrinos, acting
alone, more appreciably raise $\ye$ for $r > 25$ km.  Still, the geometric
effect dominates the extinction of the neutrino flux and so even here
$\ye$ is not radically altered outside of the optically thick region. 

\subsection{$\dot{m} \gtrsim 1$} \label{thickresults}

Both neutrinos and antineutrinos become trapped in the inner regions of
accretion disks with higher accretion rates ($\dot{m} \gtrsim 1$).  For
our calculations of $\ye$ in such disks we would prefer to use disk models
that incorporate neutrino transport effects.  We therefore switch to the
DPN disk models, which, unlike the PWF models, self-consistently include
the effects of neutrino opacity.  We find that given the PWF parameters, 
their model is optically thin,
whereas given the DPN parameters their model is optically thick.
Our results for $\ye$ in the inner
regions of these disks are very sensitive to this choice. 

\begin{figure}
\plotone{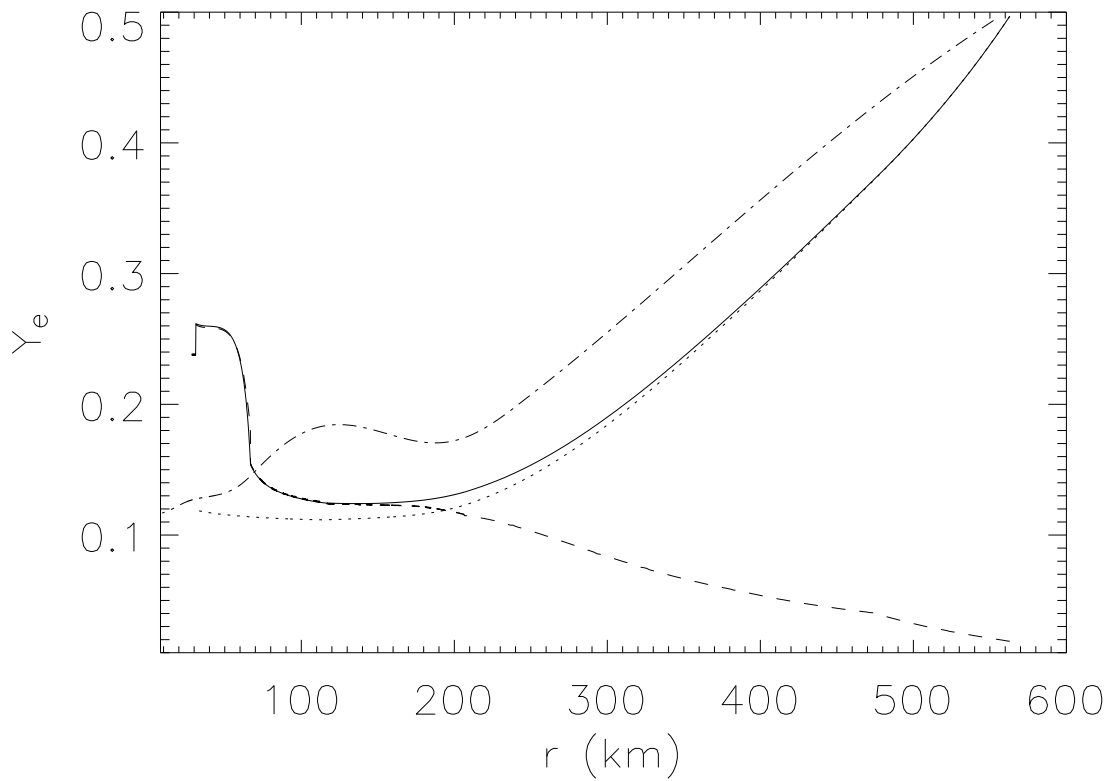}
\caption{Electron fraction $\ye$ as a function of radius for the DPN and 
PWF disk models with $\dot{m}=1$, black hole spin parameter $a=0$, and 
alpha viscosity $\alpha=0.1$.  The dot-dashed line shows $\ye$ calculated 
with the PWF disk model, while the remaining three lines are from the DPN 
disk model. The solid line shows $\ye$ from our full calculation, the dotted 
line is for a calculation without neutrino interactions, and the dashed line 
shows the equilibrium $\ye$.
\label{fig:dpn1}}
\end{figure}

Fig.~\ref{fig:dpn1} shows our calculated electron fraction for the
$\dot{m}=1$ DPN model.  Again, the solid line is our full calculation, the
dotted line is our calculation with the neutrinos removed, and the dashed
line is the equilibrium electron fraction.  Also included for comparison
is our full calculation for the PWF $\dot{m}=1$ model with $\alpha=0.1$
and $a=0$.  The temperatures and densities here are significantly higher
than for $\dot{m}=0.1$, and, as a result, electron capture quickly drives
$\ye$ to a very low value.  Positron capture moderates this drop within $r
\sim 250$ km.  As shown, neutrino interactions radically alter $\ye$ in
the DPN model.  Once neutrino trapping sets in at $r \sim 70$ km,
thermalization decreases the neutrino temperature slightly but raises the
effective flux; the latter wins out and the neutrino capture rates
increase markedly.  This drives $\ye$ back up to 0.26 before
antineutrino trapping sets in and relowers $\ye$ to $\sim 0.24$. 

As in the PWF $\dot{m}=0.1$, $a=0.95$ disk model from section
\ref{thinresults}, we see that neutrinos have a noticeable impact within
the optically thick region and much less influence outside of that region. 
Here the sections of the disk where neutrinos are at least partially
trapped are larger and so the resulting $\ye$ in the inner disk (0.25
compared to closer to 0.1 in the PWF model or the DPN model without
neutrinos) is more dramatically altered.   $\ye$ in the partially
trapped region is particularly sensitive to the neutrino physics; when we
artificially modify $T_{\nu}$ by only a few percent we find the resulting
$\ye$ in this region can change by 30\% or more. However, outside of the
trapping regions, the influence of neutrinos is limited by the same
factors: extinction, competition between neutrino and antineutrino
captures, and, most importantly, the geometric effect. 

\begin{figure}
\plotone{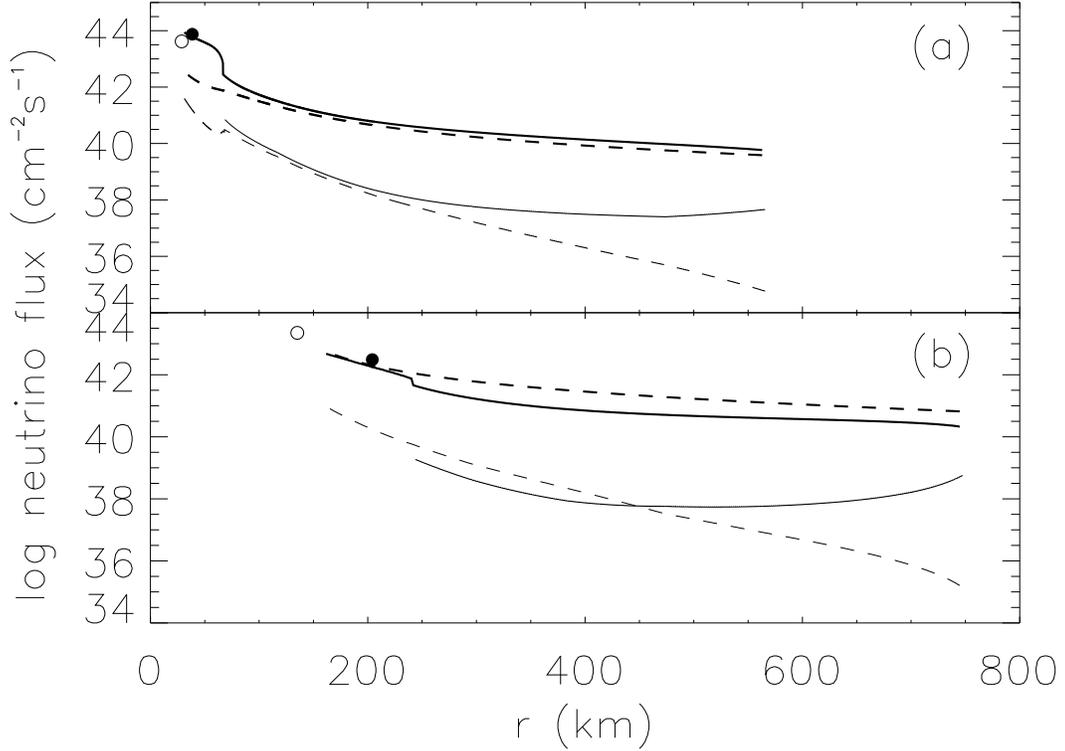}
\caption{Net neutrino fluxes versus radius for the DPN disk models with 
(a) $\dot{m}=1$ and (b) $\dot{m}=10$.  The dark solid (dashed) lines show the 
net effective (anti)neutrino flux at each radius, while the lighter weight 
lines show the net neutrino flux emitted from each optically thin zone.  The 
filled and unfilled circles depict the net antineutrino and neutrino fluxes, 
respectively, from the optically thick region and 
are plotted at the effective `neutrinosurface' radii $r_{\nu}$. 
\label{fig:netflux}}
\end{figure}

Fig.~\ref{fig:netflux}a shows the net neutrino (solid lines) and
antineutrino (dashed lines) fluxes as a function of radius for the DPN
$\dot{m}=1$ model.  The thin lines show the net fluxes emitted in the
optically thin regions, and the filled and unfilled circles show the net
neutrino and antineutrino fluxes, respectively, emitted from the optically
thick regions.  The thick lines show the calculated effective fluxes at
each radius due to contributions from the rest of the disk (except for the
`partially trapped' region, $38 \lesssim r \lesssim 68$ km, where the
effective neutrino flux is from a Fermi-Dirac distribution as described in
section \ref{partrap}).  It is important to note that in the optically
thin regions the total effective fluxes at each radius are significantly
higher than the emitted fluxes.  In these regions the effective fluxes are
dominated by contributions from the rest of the disk, particularly from
the optically thin regions interior to that zone where the emitted fluxes
are higher and from the optically thick region. Additionally, though the
emitted antineutrino flux is everywhere smaller than the neutrino flux
(and often orders of magnitude smaller), the effective antineutrino flux
is less than a factor of two smaller than the effective neutrino flux for
most of the disk with $r > 100$ km. 
 
\begin{figure}
\plotone{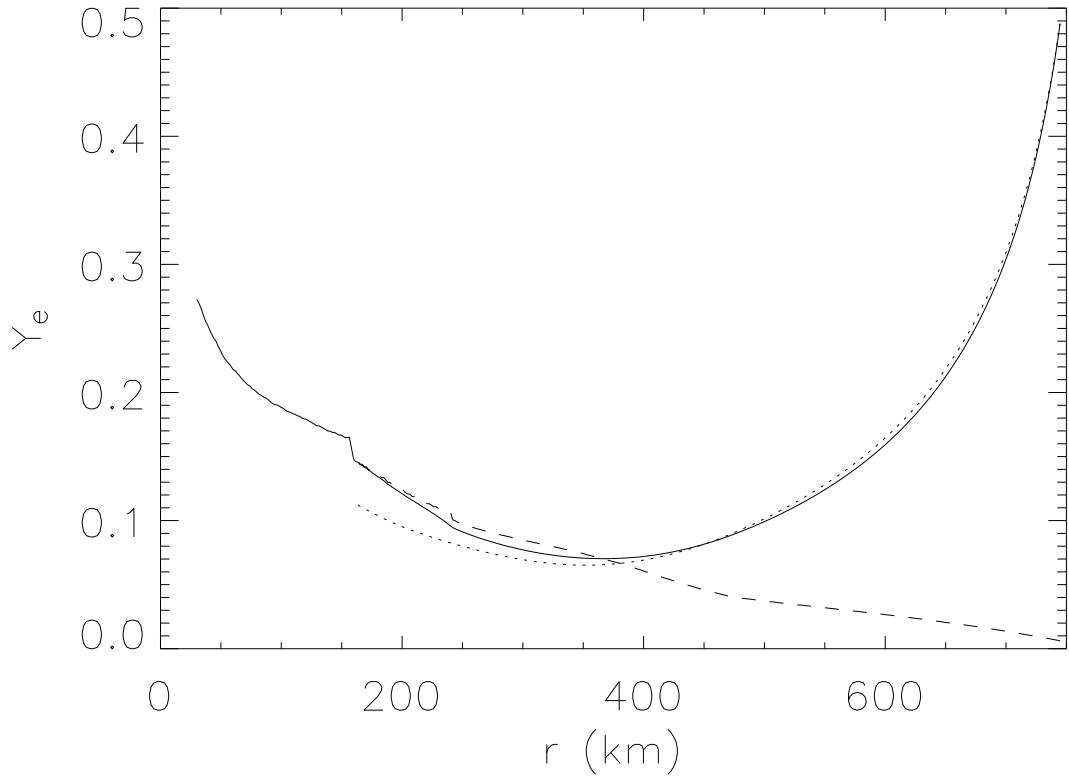}
\caption{Electron fraction $\ye$ as a function of radius for the DPN disk 
model with $\dot{m}=10$, $a=0$, and $\alpha=0.1$.  As in Fig.~\ref{fig:dpn1}, 
the solid line is the full calculation with neutrino interactions, the 
dotted line without, and the dashed line shows the equilibrium $\ye$.
\label{fig:dpn10}}
\end{figure}

Fig.~\ref{fig:dpn10} shows our calculated electron fraction for the
$\dot{m}=10$ DPN model.  As in Fig.~\ref{fig:dpn1}, the solid line is our
full calculation, the dotted line is our calculation with the neutrinos
removed (for the optically thin and partial trapping zones only), and the
dashed line is the equilibrium $\ye$.  The evolution of $\ye$ here is
similar to the $\dot{m}=1$ case, with one notable exception.  The inner
regions of the disk are much hotter than the $\dot{m}=1$ model, and so
positron capture plays a much larger role.  This is indicated in the
calculation without neutrinos, where $\ye$ begins to increase at $r \sim
300$ km.  The greatest influence, though, is in the optically thick region
($r \lesssim 160$ km).  Here $\ye$ continues to increase, albeit at a
slower rate, even when the antineutrinos are trapped.  Since $Y_{L}$ is
(approximately) constant in this region, $Y_{\nu}$ becomes negative, which
favors antineutrinos over neutrinos.  As a result the antineutrino flux
from the optically thick region is significantly higher and more energetic
than the neutrino flux. 

The emitted and effective neutrino fluxes in this model are shown in
Fig.~\ref{fig:netflux}b, where the lines and points are defined as in
Fig.~\ref{fig:netflux}a.  Again, the major difference between the two
panels is the higher antineutrino flux from the optically thick region. 
This carries over to the outer regions of the disk, where the effective
antineutrino flux is approximately three times that of the neutrino flux. 
Still, this difference isn't large, and the effective flux suffers from
the same geometric effect as the other models.  As a result, the influence
of the neutrinos within the disk is again largely confined to the regions
in and around where they are trapped. 

\begin{figure}
\plotone{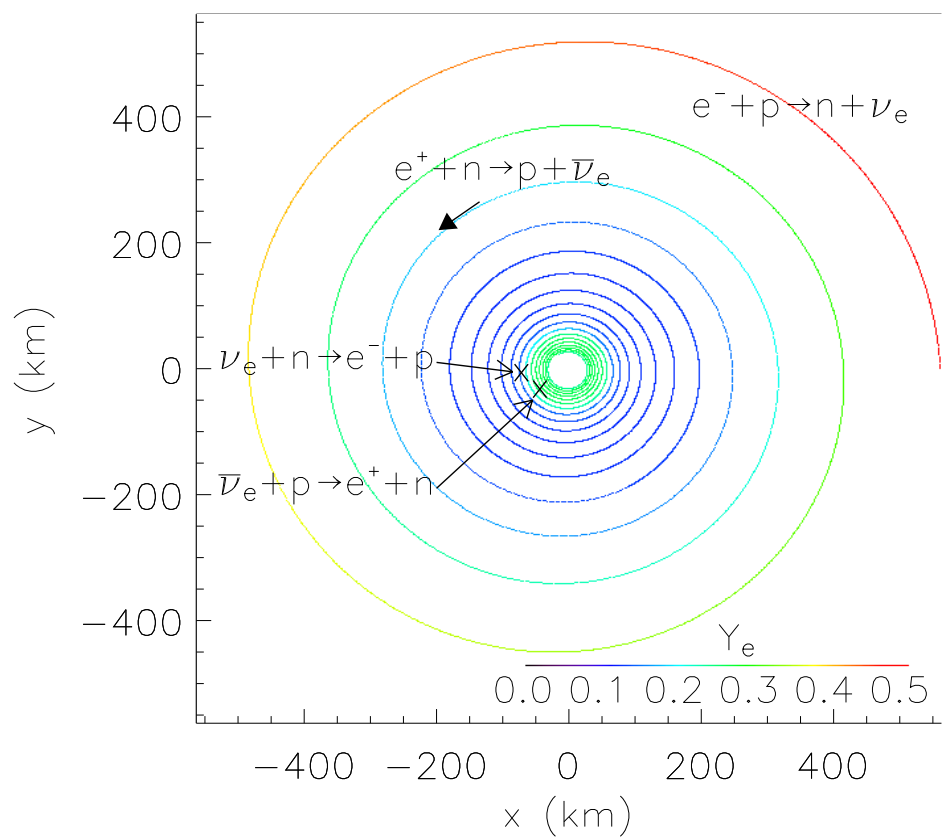}
\caption{Electron fraction $\ye$ as a function of disk position $(x,y)$
for an inspiraling mass element in the DPN disk model with $\dot{m}=1.0$,
$a=0$, and $\alpha=0.1$.  The markings indicate where each capture
reaction becomes important, defined as where the absence of that reaction
in the calculation affects $\ye$ by greater than $10\%$. 
\label{fig:cplot1dpn}}
\end{figure}

\begin{figure}
\plotone{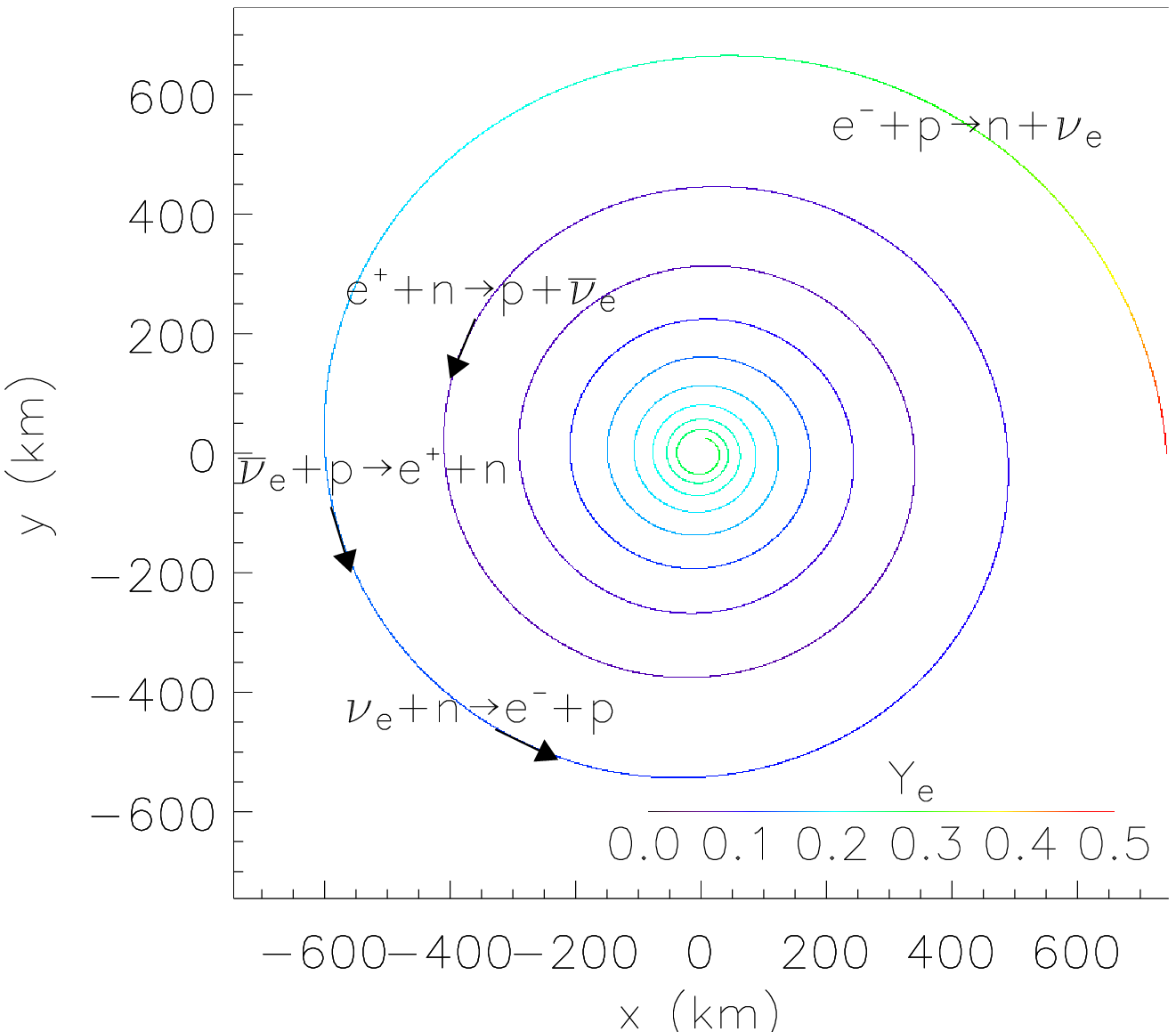}
\caption{Same as Fig.~\ref{fig:cplot1dpn} for DPN disk model $\dot{m}=10$.
\label{fig:cplot10dpn}}
\end{figure}

The results of this section are summarized in Figs.~\ref{fig:cplot1dpn} and
\ref{fig:cplot10dpn}.  These figures show the electron fraction $\ye$ as a
function of position for inspiraling mass elements in the DPN disk models
$\dot{m}=1$ and $10$, respectively.  We note the importance of positron,
neutrino, and antineutrino capture by marking the points where each
interaction, when omitted, changes $\ye$ by more than $10\%$. 

We have calculated the fluxes coming everywhere from the surface 
of the disk and they affect many important pieces of physics.  They
alter the electron fraction in material which flows off the disk
and they also determine the neutrino-antineutrino annhilation rate 
above the disk. A preliminary analysis of the former is in the next section.
An analysis of the latter will be done in future work, but
to give a point of reference: 
for the DPN model with $\dot{m} = 1$, 
along the z-axis, at a height
of $8 \times 10^6 {\rm cm}$ (again neglecting relativistic corrections)
the neutrino number flux is $1.27 \times 10^{42} {\rm cm}^{-2}$,
the antineutrino number flux is $5.0 \times 10^{41} {\rm cm}^{-2}$.
The average energies of the neutrinos and antineutrinos at this point
are 15.1 MeV and 15.7 MeV respectively, where as the rms energies are
16.4 MeV and 17.1 MeV respectively.

\section{Preliminary Outflow Model} \label{otflmod}

We further examine the influence of neutrinos on nucleosynthesis in GRB's
by following material from the disk as it is ejected in a wind.  Neutrinos
leaving the disk will interact with the outflow material, thus changing
its electron fraction.  Our goal here is not to develop a realistic
outflow model but to estimate the import of neutrino interactions on
nucleosynthesis in the wind.  To this end we choose a simple constant
velocity, adiabatic outflow.  We assume a constant mass outflow rate
proportional to $4\pi r^{2} \rho v$, which gives $\rho \sim r^{-2}$, where
$r$ here is the outflow radius in the spherical approximation.  The
temperature in the ejecta is found from the density and the entropy
per baryon.  Here we assume a constant entropy per baryon in the outflow.
The entropy per baryon includes the contributions from relativistic
particles and nucleons, as in \citet{qia96}: 
\begin{equation} 
s/k_{b} \approx 0.052 \frac{T_{\mathrm{MeV}}^3}{\rho_{10}} + 7.4 + \ln
\left(\frac{T_{\mathrm{MeV}}^{3/2}}{\rho_{10}} \right), 
\label{eq:entropy}
\end{equation} 
where $T_{\mathrm{MeV}}$ is the temperature in MeV and $\rho_{10}$ is the
density in units of $10^{10}$ g/cm$^{3}$.  We start with a mass element in
the disk, launch it with velocity $v$ equal to the escape velocity, and
follow the evolution of the electron fraction in the ejecta as described
in section~\ref{opthin} until the temperature drops to $10^{10}$ K. 

An important difference between this calculation and that of
section~\ref{opthin} is the evaluation of the effective neutrino and
antineutrino fluxes at each point.  We simplify the evaluation of the
effective fluxes for points above the disk by assuming that the disk is
essentially flat as viewed from above.  Therefore each disk zone is no
longer a cylindrical shell but a flat ring with radius $r_{j}$ and width
$dr_{j}$.  The emitted fluxes from each region first need to be adjusted
for this change in geometry.  Instead of multiplying the number of
neutrinos emitted per second per keV per volume ($\sigma_{e} c f_{e}$ from
Eqn.~\ref{eq:nuj}) by the width $dr_{j}$ of the zone for a cylindrical
emitting surface, we multiply this quantity by the depth $d_{j}$ of the
emitting zone.  For the totally optically thin regions, the depth is just
the total height of the disk, $d_{j}=2H$.  For zones where the neutrino
mean free path $l_{\nu}$ is less than $2H$, we have $d_{j}=l_{\nu}$.  The
height correction $h_{j}$ discussed in section~\ref{opthin} is also
applicable here.  The emitted neutrino flux $\phi_{\nu,j}^{\prime}$ from
each zone $j$ is therefore
\begin{equation}
\phi_{\nu,j}^{\prime}(E_{\nu}) = \phi_{\nu,j}(E_{\nu}) \times h_{j} 
\frac{d_{j}}{dr_{j}}. 
\label{eq:nujout}
\end{equation}
An equivalent expression holds for the antineutrinos.

The emitted fluxes from the optically thick region are found using the
`neutrinosurfaces' calculated as described in section~\ref{opthick}.  At
each radius within the optically thick region, the emitted fluxes are
calculated from Fermi-Dirac distributions according to
Eqn.~\ref{eq:FDflux}.  The temperature in this expression is taken to be
equal to the vertical disk temperature at the appropriate decoupling
height $h_{\nu}$ or $h_{\bar{\nu}}$ for that radius, and the chemical
potential is equal to $\mu_{\nu}$ or $\mu_{\bar{\nu}}$ at that disk
radius.  These temperatures vary along the 
'neutrinosurface'.  For example for DPN 
$\dot{m} = 10$, $ 2.4 \lesssim T_\nu \lesssim 6.0$ and $3.6 \lesssim
T_{\bar{\nu}} \lesssim 6.8$, while for DPN $\dot{m} = 1$, 
$ 4.5 \lesssim T_\nu \lesssim 5.3$ and $5.1 \lesssim
T_{\bar{\nu}} \lesssim 5.4$.

\begin{figure}
\plotone{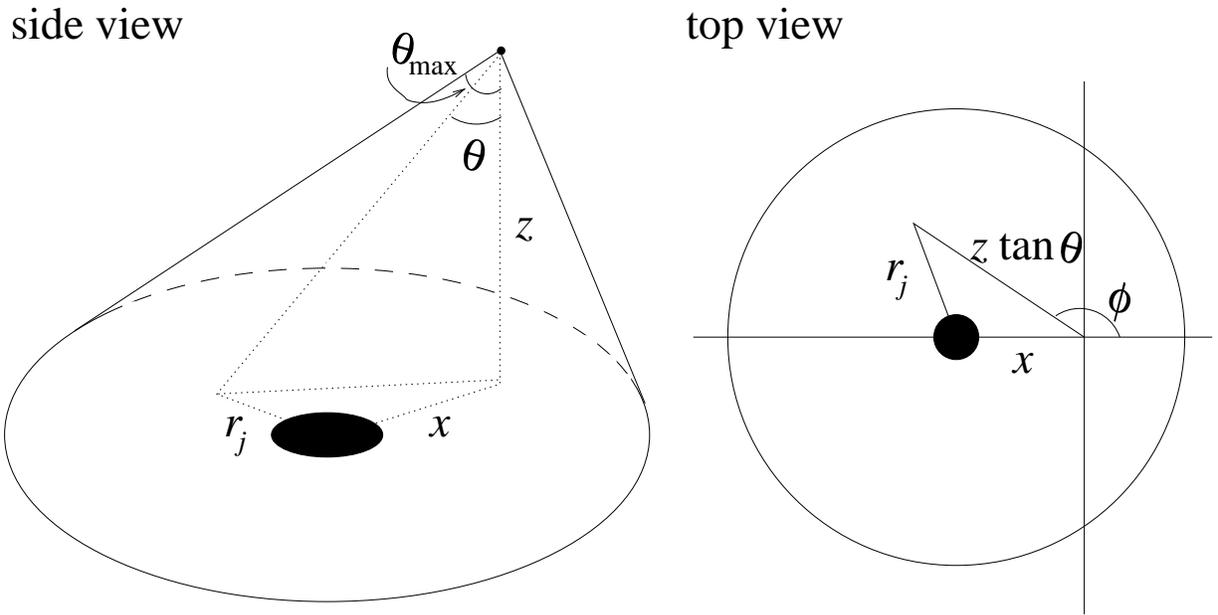}
\caption{Illustration of the angular geometry relevant to calculating the 
effective neutrino flux for a point $(x,y,z)$ above the disk.
\label{fig:disk}}
\end{figure}

The effective fluxes are found by integrating the emitted fluxes over the
entire disk, similar to in section~\ref{opthin} but with very different
geometry. Here the appropriate solid angle is not so easily decoupled from
the emitted fluxes, so we evaluate the full integral for the effective
flux $\phi_{\nu}^{\prime\mathrm{eff}}$ at each point $(x,y,z)$ in the
outflow,
\begin{equation}
\phi_{\nu}^{\prime\mathrm{eff}}(E_{\nu}) = \int_{0}^{2\pi} 
\int_{0}^{\theta_{\mathrm{max}}(\phi)} \phi_{\nu}^{\prime}(E_{\nu},\theta,\phi)
\frac{\sin \theta d\theta d\phi}{4\pi}.
\label{eq:effluxout}
\end{equation}
In the above expression, $\phi_{\nu}^{\prime}(E_{\nu},\theta,\phi)$ is the
emitted flux $\phi_{\nu,j}^{\prime}$ at $r_{j} = [z^{2}\tan^{2}\theta +
x^{2} + 2xz\tan \theta \cos \phi]^{1/2}$, and
$\theta_{\mathrm{max}}(\phi)$ is the maximum value for $\theta$ at the
disk edge.  This angular geometry is illustrated in Fig.~\ref{fig:disk}. 

\begin{figure}
\plotone{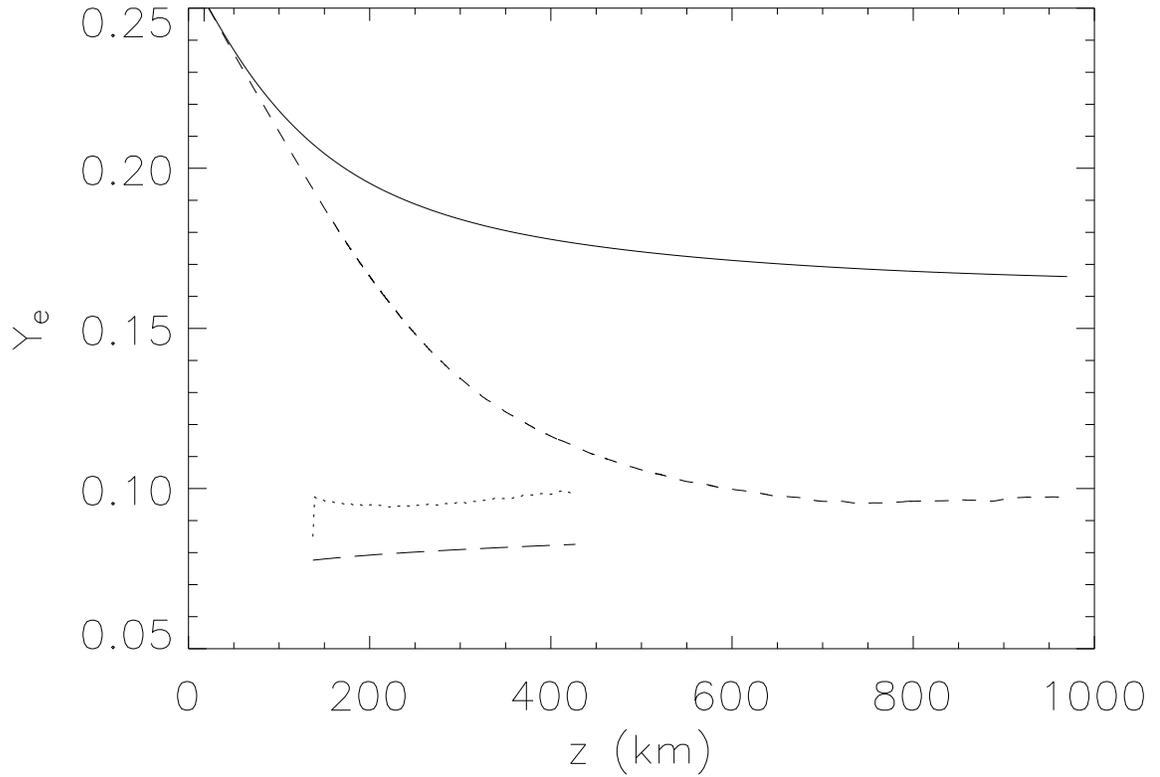}
\caption{Electron fraction versus height in the outflow originating from two 
locations within the DPN disk model $\dot{m}=10$: close to the black hole 
($r \sim 35$ km, solid line) and at the outer edge of the optically thick 
region ($r \sim 250$ km, long-dashed line).  The short-dashed line and dotted 
line show the equilibrium $\ye$ for each outflow. 
\label{fig:yeoutflow}}
\end{figure}

Fig.~\ref{fig:yeoutflow} shows the minimum variation in $\ye$ for two such
calculations for outflows from the DPN $\dot{m}=10$ disk model.  The solid
line is for an outflow from close to the black hole, at $r_{\mathrm{disk}}
\sim 35$ km.  The short-dashed line is the equilibrium $\ye$ for the same
outflow.  Electron capture and antineutrino capture here combine to lower
$\ye$ from its initial value in the disk.   As the electron capture
rate drops, the equilibrium $\ye$ (meaning that $\ye$ that would obtain if
the weak rates were in equilibrium) is increasingly set by neutrino and
antineutrino capture.  However, $\ye$ itself levels off before then since
the very rapid outflow velocity ($v=v_{\mathrm{esc}}=1.5\times 10^{10}$
cm/s) causes $\ye$ to quickly fall out of equilibrium. The second outflow
example, the long-dashed line in Fig.~\ref{fig:yeoutflow}, starts from the
disk region just outside of the optically thick zone and immediately falls
out of equilibrium.  Here the reason is not so much that the outflow
velocity is high, but that the equilibrium $\ye$ rises sharply above
the disk.  This is entirely a neutrino-driven effect.  Within the disk,
the neutrino and antineutrino fluxes are quickly diminished by extinction
and, most importantly, the geometric effect, so the equilibrium $\ye$ is
set largely by electron and positron capture.  Above the disk, however,
the outflow material is exposed to the high neutrino and antineutrino
fluxes leaving the disk, and so the jump in the neutrino and antineutrino
capture rates readjusts the equilibrium $\ye$, here to a much higher
value.  In this example, the temperature falls below $10^{10}$ K before
$\ye$ levels off, but still the outflow velocity is too great (here
$v=v_{\mathrm{esc}}=5.7\times 10^{9}$ cm/s) for equilibrium to be
established. 

\begin{figure}
\plotone{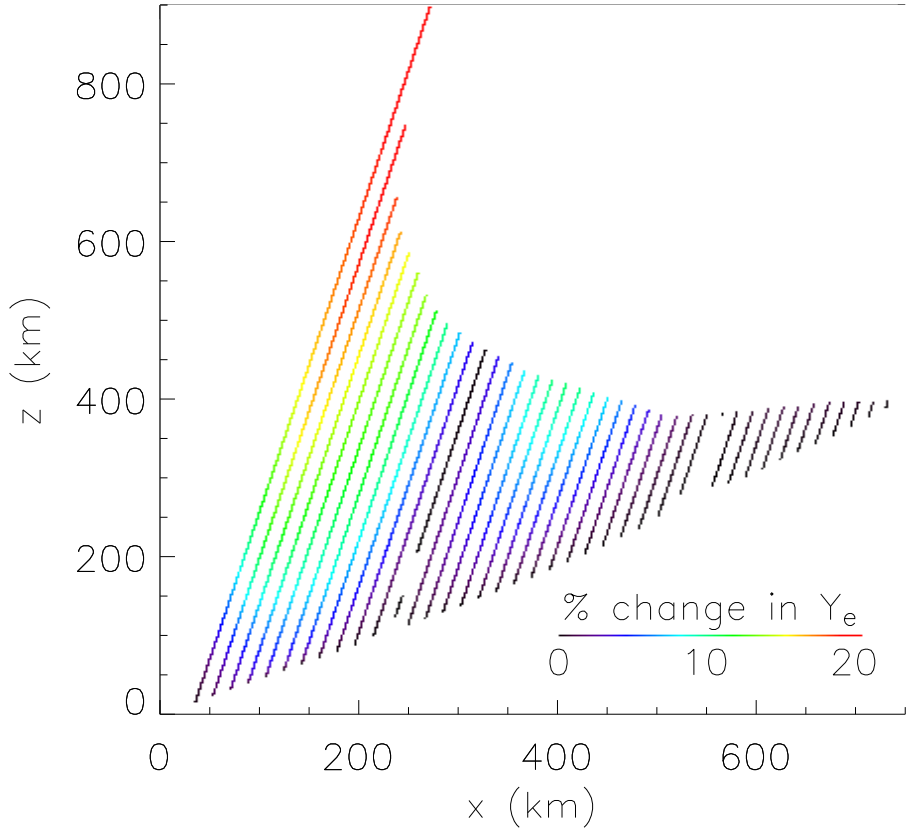}
\caption{Percent change in $\ye$, $(\ye^{\mathrm{with}\;\nu} -
\ye^{\mathrm{no}\;\nu})\times 100/\ye^{\mathrm{no}\;\nu}$, is plotted for
outflows from DPN disk model $\dot{m}=10$.  $\ye^{\mathrm{with}\;\nu}$
includes the effects of neutrino interactions, while $\ye^{\mathrm{no}\;\nu}$ 
is calculated with the neutrino and antineutrino capture rates set
to zero. 
\label{fig:yecomp}}
\end{figure}

\begin{figure}
\plotone{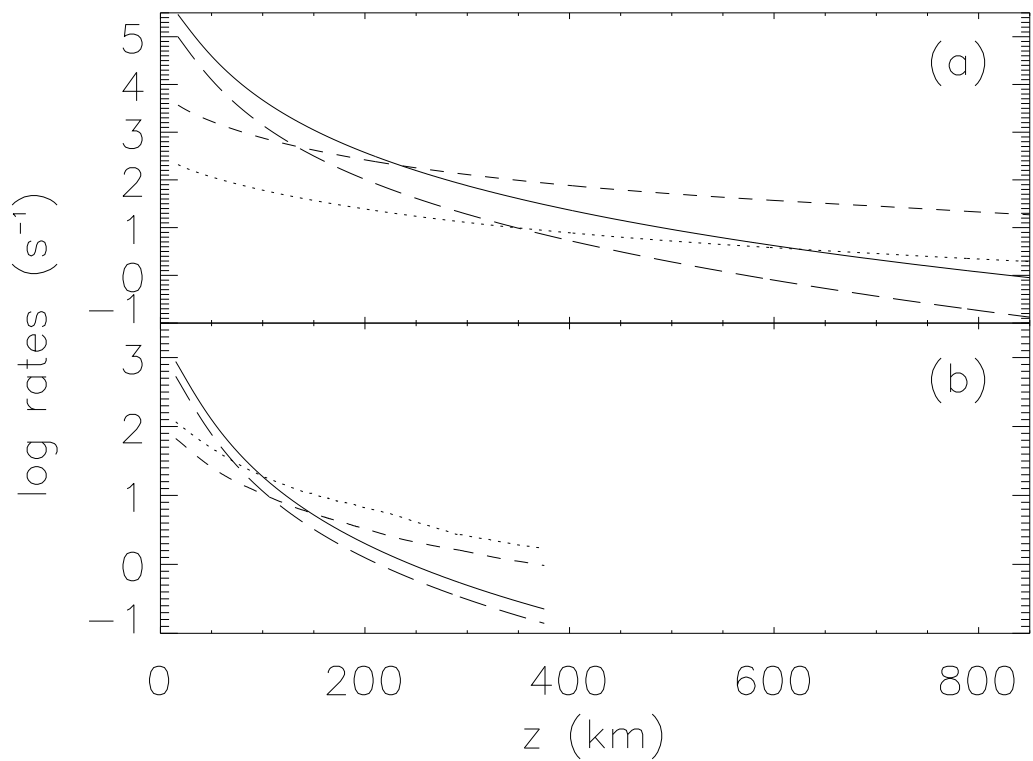}
\caption{Capture rates versus height in the outflow originating from two
disk models, (a) DPN $\dot{m}=10$ and (b) PWF $\dot{m}=0.1$, $a=0.95$. 
For both models, the solid line is the electron capture rate, the
long-dashed line is the positron capture rate, the short-dashed line is
the antineutrino capture rate, and the dotted line is the neutrino capture
rate.  As shown, the neutrino and antineutrino capture rates remain high
well above the disk. 
\label{fig:rates}}
\end{figure}

Figs.~\ref{fig:yecomp} and \ref{fig:rates} further illustrate the
importance of neutrino capture in the wind.  Fig.~\ref{fig:yecomp}
compares the outflow $\ye$ calculated as above to that calculated without
including neutrino and antineutrino captures. The percent change in $\ye$,
$(\ye^{\mathrm{with}\;\nu} - \ye^{\mathrm{no}\;\nu})\times
100 \% /\ye^{\mathrm{no}\;\nu}$, is plotted for outflows from every ten radial
zones in the DPN disk model with $\dot{m}=10$.  In outflows from the outer
disk, the neutrino flux is small and so is the percent change in $\ye$. 
Above the inner disk, particularly above the optically thick region,  
the neutrino flux is much higher and $\ye$ is influenced accordingly.
The reason the neutrinos can have such a large influence
is shown in Fig.~\ref{fig:rates}.  It plots the capture rates in two
outflow models, (a) DPN $\dot{m}=10$ and (b) PWF $\dot{m}=0.1$, $a=0.95$,
as a function of height.  In both cases, the electron and positron capture
rates fall quickly, dropping below the neutrino and antineutrino rates
within a few hundred kilometers.  The neutrino rates drop much more slowly
due to the geometry of the disk.  If the disk emitted neutrinos uniformly
and the outflow remained above the disk, the rates would be almost
constant. In our physical disks, this doesn't hold exactly since most of
the neutrinos are from the inner regions of the disk, but still the rates
fall much more slowly than, for example, $\sim r^{-2}$ as in the
proto-neutron star case. 

It is important to note that our outflow model only conservatively estimates 
the influence of neutrino interactions in the wind.  For example, if the 
velocity of the outflow is at any point slower than the escape velocity at 
the starting point of the flow, the wind's exposure to the neutrino flux will 
be correspondingly greater.  In addition, we only follow the outflow until 
the temperature drops below $10^{10}$ K; we expect neutrino interactions 
to continue to play an important role in the subsequent nucleosynthesis.  
This is true even for disks that are almost entirely optically thin (see, 
for example, Fig.~\ref{fig:rates}b).

\section{Conclusions} \label{concl}

Here we have made a detailed analysis of the importance of the weak rates
in the disks of gamma ray bursts.  We follow mass elements as they spiral
into the center of these disks, keeping track of all the rates, the
emitted neutrino and antineutrino fluxes, and the electron fraction. We
have included not only the effect of electron and positron capture but
also of neutrino and antineutrino capture.  Neutrino and antineutrino
capture play an essential role for models with high accretion rates, where
the neutrinos become trapped.  Here the electron fraction can change by
factors of several over calculations where these rates are not included. 
In addition, neutrinos also play an important role in models with lower
accretion rates if they have high black hole spin parameter. 

Even if the weak rates come into equilibrium toward the center of the disk, 
it is still necessary to follow the complete evolution of the mass element as
we have done.  This is because the lepton number of the material as it falls 
into the optically thick regions goes into determining the electron fraction
and fluxes of neutrinos in these areas.

As part of this study, we have calculated the 'neutrinosurfaces' or surfaces 
that define the regions that are optically think to neutrinos.  We find
that these are not spheres as in the protoneutron star case, but take on
different shapes, usually more wedgelike.

As part of the larger picture,  we wish to determine the nucleosynthesis that
will come from the mass outflow from these disks.  This will involve a 
hydrodynamic calculation, in addition to simply knowing the
$Y_e$ on the disk which we have calculated here.  However, whatever this
outflow looks like, it is certain that the neutrino and antineutrino
flux from the disk will hit it from behind as it moves out.  This will
change the electron fraction and therefore also any calculation
of nucleosynthesis.  We have made an estimate of the minimum influence of
this effect by using a constant velocity for the outflowing material 
which is equal to the escape velocity.  Even in this case, the neutrinos 
change the $Y_e$ in the outflow by 10\%-60\%.

Future studies of nucleosynthesis from these disks will produce exciting 
results.  Whatever the results will be, 
it is certain that the neutrino interactions
will have to be included in the calculations.

\acknowledgments

\appendix

\section{Appendix}

\begin{figure}
\plotone{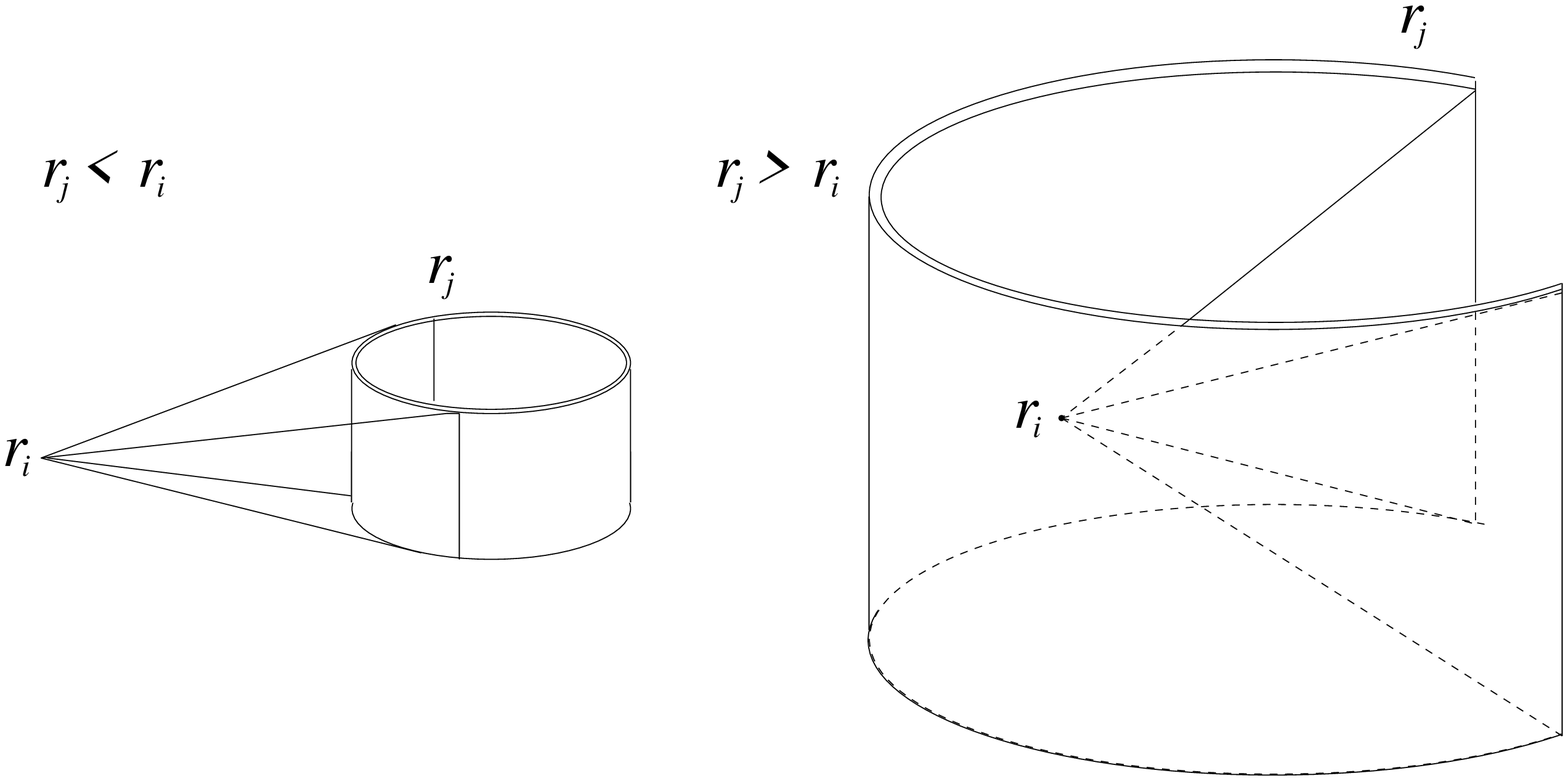}
\caption{Illustration of the solid angle, $d\Omega_{i,j}$, subtended by 
zone $j$ as viewed from zone $i$, for where $j$ is interior to $i$, i.e. 
$r_{j} < r_{i}$, and where $j$ is exterior to $i$, i.e.  
$r_{j} > r_{i}$.  Note that for $r_{j} > r_{i}$, part of zone $j$ is 
blocked by the black hole and (possibly) the optically thick region. 
We calculate the flux at the point $r_{i}$ from every accessible place 
on the surface at $r_{j}$.
\label{fig:solidang}}
\end{figure}

Here we describe the evaluation of the integral over the solid angle $\int
d\Omega_{i,j}$ from Eqn.~\ref{eq:totalflux}, where $d\Omega_{i,j}$ is the
solid angle subtended by zone $j$ as viewed from zone $i$ within the disk. As depicted in Fig.~\ref{fig:solidang}, we take zone $j$ to be a uniform
cylindrical shell with radius $r_{j}$, height $2H_{j}$, and thickness
$dr_{j}$ and the viewpoint in zone $i$ to be a point at the midplane of
the disk a distance $r_{i}$ from the black hole.  The general form of the
integral is as follows: 
\begin{equation}
\int d\Omega_{i,j} = \int^{\phi_{b}}_{\phi_{a}} 
\int^{\theta_{b}(\phi)}_{\theta_{a}(\phi)} \sin \theta d\theta d\phi.
\label{eq:inte}
\end{equation}
The solid angle geometry is dependent upon whether zone $j$ is interior or
exterior to zone $i$, and so the limits of integration for each case are
treated separately below.  Once the limits are determined we solve the
integrals in Eqn.~\ref{eq:inte} numerically using the extended Simpson's rule.

\begin{figure}
\plotone{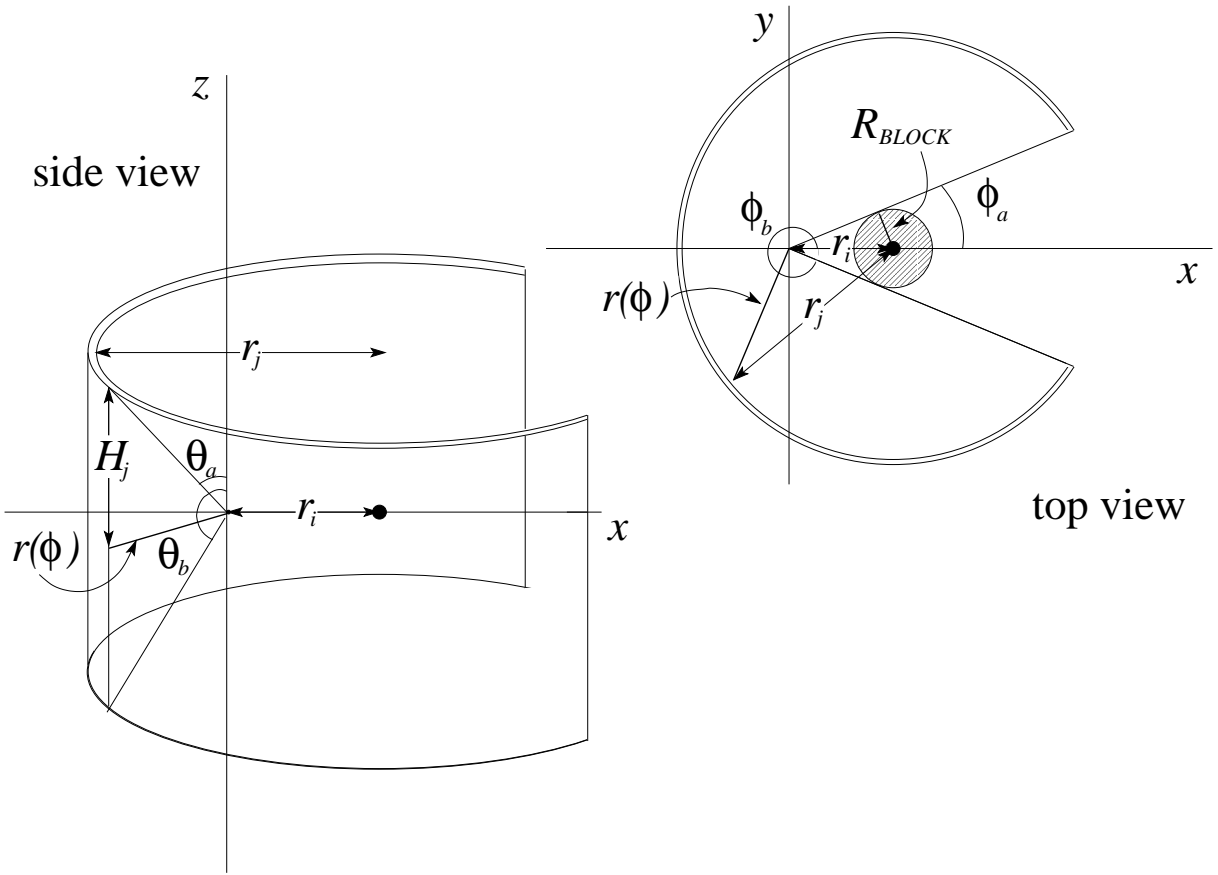}
\caption{More detailed illustration of the solid angle, $d\Omega_{i,j}$,
subtended by zone $j$ where $j$ is exterior to $i$, i.e. $r_{j} > r_{i}$. 
\label{fig:saji}}
\end{figure}

The case where zone $j$ is exterior to zone $i$ is pictured in more detail
in Fig.~\ref{fig:saji}.  For convenience we arrange the coordinate axes as
shown, with the origin at the viewpoint in zone $i$, the $+x$ axis running
along the midplane of the disk and through the black hole, and the $z$
axis extending vertically above and below the disk.  In every case where
$r_{j}>r_{i}$, part of zone $j$ is blocked by the event horizon and
possibly an optically thick region, as depicted in the top view detail of
Fig.~\ref{fig:saji}.  Therefore the limits of the integral over the $\phi$
coordinate become
\begin{eqnarray}
\phi_{a} & = & \sin^{-1} \left( \frac{R_{BLOCK}}{r_{i}} \right)
\label{eq:p1a} \\
\phi_{b} & = & 2\pi - 
\sin^{-1} \left( \frac{R_{BLOCK}}{r_{i}} \right), \label{eq:p1b}
\end{eqnarray}
where $R_{BLOCK}$ is the Schwarzschild radius or, if present, the radius of
the optically thick region. 

The limits of the integral over $\theta$, as illustrated in the side view
detail of Fig.~\ref{fig:saji}, are
\begin{eqnarray}
\theta_{a}(\phi) & = & \tan^{-1} \left( \frac{r(\phi)}{H_{j}} \right) 
\label{eq:t1a} \\
\theta_{b}(\phi) & = & \pi - \theta_{a}(\phi), \label{eq:t1b}
\end{eqnarray}
where $r(\phi)$ is the distance from the origin to the midplane of zone
$j$ as a function of $\phi$.  The distance $r(\phi)$ is found from the law of 
cosines 
\begin{equation}
r_{j}^{2} = r_{i}^{2} + r^{2} - 2 r_{i} r \cos \phi
\label{eq:cosines}
\end{equation}
to be
\begin{equation}
r(\phi) = r_{i} \cos \phi + (r_{j}^{2} - r_{i}^{2} \sin^{2} \phi)^{1/2}.
\label{eq:rphi}
\end{equation}

\begin{figure}
\plotone{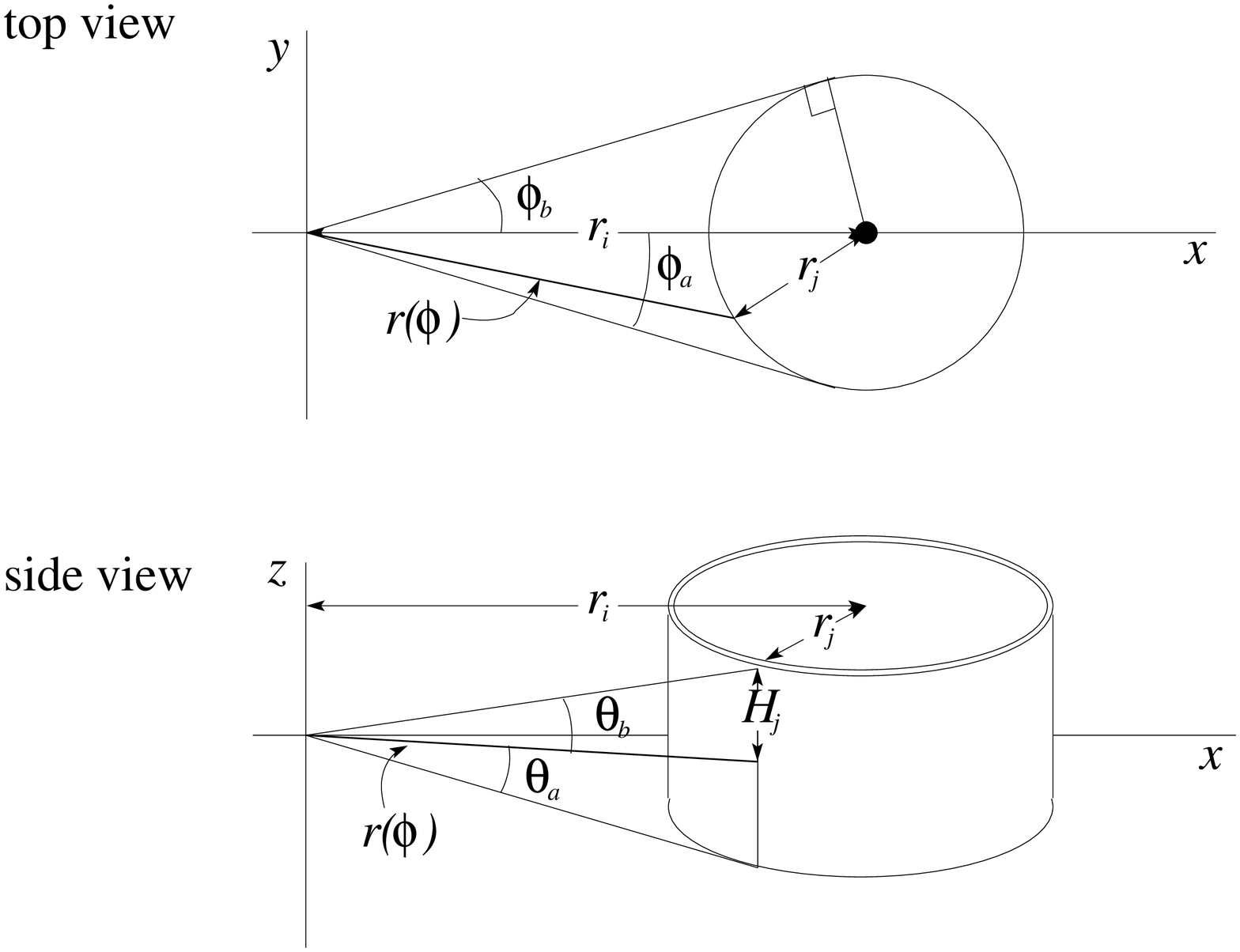}
\caption{More detailed illustration of the solid angle, $d\Omega_{i,j}$,
subtended by zone $j$ as viewed from zone $i$, for where $j$ is interior
to $i$, i.e.  $r_{j} < r_{i}$. 
\label{fig:saij}}
\end{figure}

The case where zone $j$ is interior to zone $i$ is illustrated in
Fig.~\ref{fig:saij}.  The limits of the $\phi$ integral, as shown in the 
top view detail, are
\begin{eqnarray}
\phi_{a} & = & - \sin^{-1} \left( \frac{r_{j}}{r_{i}} \right) \label{eq:p2a} \\
\phi_{b} & = & + \sin^{-1} \left( \frac{r_{j}}{r_{i}} \right). \label{eq:p2b}
\end{eqnarray}
The side view detail of Fig.~\ref{fig:saij} shows the limits of the
integral over $\theta$.  The expressions for $\theta_{a}(\phi)$ and
$\theta_{b}(\phi)$ are identical to Eqns.~\ref{eq:t1a} and \ref{eq:t1b},
except here the distance $r(\phi)$ is the negative solution of 
Eqn.~\ref{eq:cosines}, 
\begin{equation}
r(\phi) = r_{i} \cos \phi - (r_{j}^{2} - r_{i}^{2} \sin^{2} \phi)^{1/2}.
\end{equation}

\acknowledgments
We wish to thank A. MacFadyen, J. Pruet and G. Raffelt for helpful
discussions.  GCM acknowledges support from the U.S. Department
of Energy under grant DE-FG02-02ER41216.




\clearpage


\begin{deluxetable}{ccccc} 
\tablewidth{0pt}
\tablecaption{The Electron Fraction $Y_{e}$ near the Event Horizon}
\tablehead{
\colhead{Model} & \colhead{$\dot{m}$} & \colhead{$a$} & \colhead{$\alpha$} 
& \colhead{final $Y_{e}$} }
\startdata
PWF & 0.1 & 0 & 0.1 & 0.45 \\
PWF & 0.1 & 0 & 0.01 & 0.05 \\
PWF & 0.1 & 0.5 & 0.1 & 0.44 \\
PWF & 0.1 & 0.95 & 0.1 & 0.14 \\
PWF & 1.0 & 0 & 0.1 & 0.12 \\
DPN & 1.0 & 0 & 0.1 & 0.24 \\
DPN & 10 & 0 & 0.1 & 0.28 \\
\enddata
\end{deluxetable}

\clearpage



\end{document}